\documentclass[aps,prl,superscriptaddress,twocolumn,amsmath,longbibliography]{revtex4-2}

\usepackage{graphicx, color}
\usepackage{xcolor}
\usepackage{amsfonts,amssymb,amsmath}
\usepackage{siunitx}
\usepackage{bm}
\usepackage{bbold}
\usepackage{braket}
\usepackage{mathtools}
\usepackage{dsfont}
\usepackage{lineno}
\usepackage{bigints}
\usepackage[normalem]{ulem}
\usepackage{enumitem}
\usepackage{amsfonts}

\begin{document}

\title{Magnetic-field-induced nonlocal transport in the topological semimetal ZrTe$_5$}

\author{Yongjian Wang}
\email[]{ywang@ph2.uni-koeln.de}
\affiliation{Physics Institute II, University of Cologne, Z\"ulpicher Str. 77, 50937 K\"oln, Germany}

\author{A. A. Taskin}
\affiliation{Physics Institute II, University of Cologne, Z\"ulpicher Str. 77, 50937 K\"oln, Germany}

\author{Yoichi Ando}
\email[]{ando@ph2.uni-koeln.de}
\affiliation{Physics Institute II, University of Cologne, Z\"ulpicher Str. 77, 50937 K\"oln, Germany}

\begin{abstract}
Nonlocal transport, which goes beyond the Ohm’s law, can be a key in understanding systems with topological order or edge states. Here we report an unusual nonlocal charge transport in the nodal-line semimetal ZrTe$_5$ that occurs in the ultra-quantum limit driven by the magnetic field applied along the $a$-axis. Surprisingly, the observed decay length of the nonlocality exceeds 100 $\mu$m and it increases linearly with the sample width. This nonlocal transport is detected not only in the longitudinal configuration, but also in the transverse one as an unusual nonlocal Hall effect.
Our findings demonstrate that the nonlocal response can offer unprecedented insights into topological quantum materials.

\end{abstract}
\maketitle

\begin{figure}[t]
	\centering
	\includegraphics[width=8.5cm]{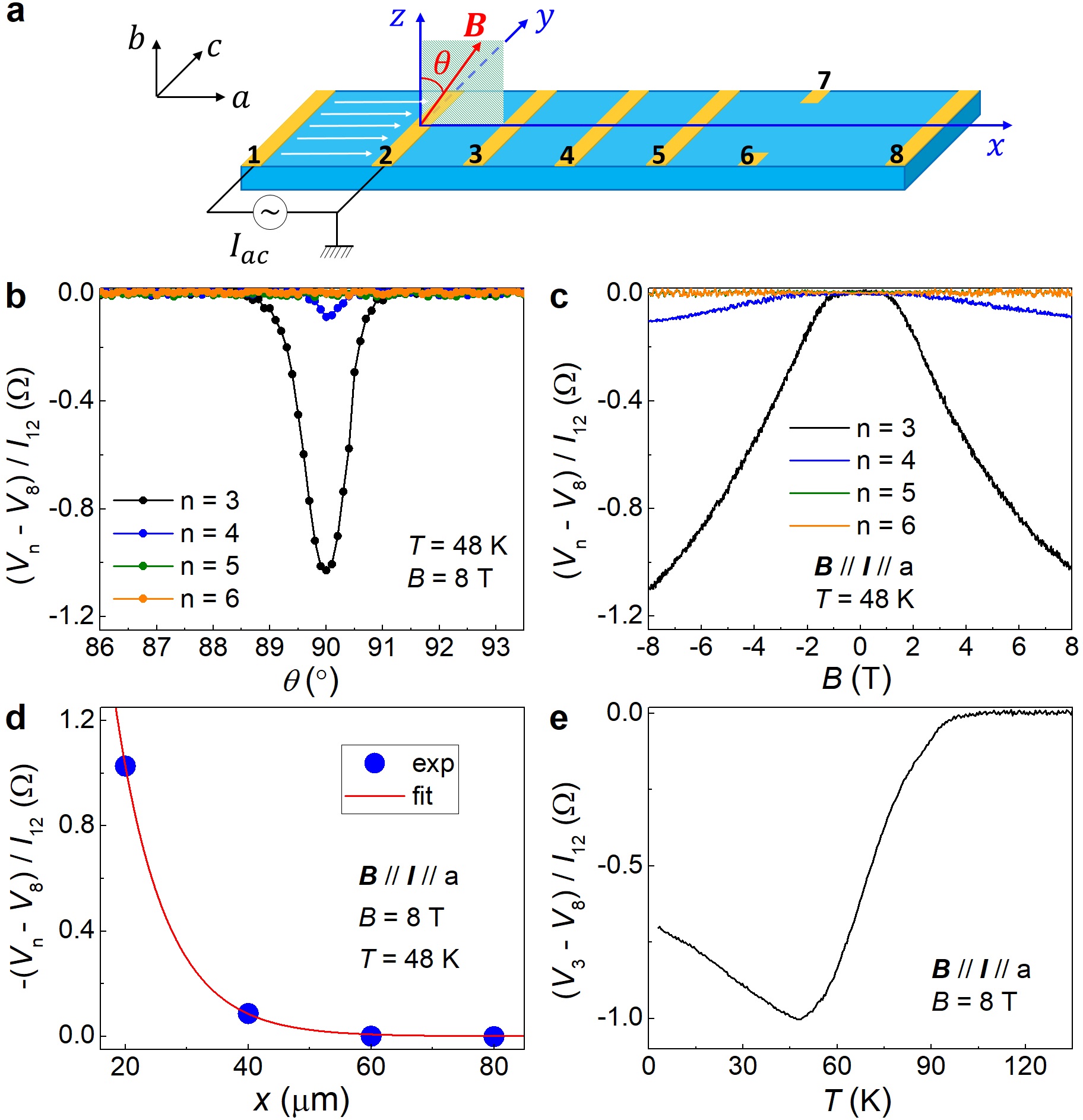}
	\caption{{\bf Nonlocal resistance $R^{\rm NL}_n$ in sample D1.} {\bf a}, Schematic drawing of the measurement configuration. Current $I_{12}$ was injected from electrode 1 and drained through the grounded electrode 2. The magnetic field $B$ was applied in the $xz$ plane with an angle $\theta$ measured from the $z$-axis. The $xyz$ transport axes correspond to $acb$ crystal axes. The $x$ coordinate is measured from the center of electrode 2. For the electrodes $n = 2-6$, the $x$ distance of 20 $\mu$m was made to be equal. {\bf b}, $\theta$-dependence of $R^{\rm NL}_n$ in 8 T at 48 K for $n = 3-6$ (electrodes 6 and 7 were shorted in this measurement). {\bf c}, $B$-dependence of $R^{\rm NL}_n$ for $n = 3-6$ measured in $B \parallel I \parallel a$-axis ($\theta$ = 90$^\circ$) at 48 K. {\bf d}, $x$-dependence of $-R^{\rm NL}_n$ at 48 K and 8 T for $B \parallel I \parallel a$-axis; the $x$ value is the coordinate position of the electrode $n$. Red line is a fit to the formula $R^{\rm NL}_n = R_0 e^{-x/\lambda}$, which gives the decay length $\lambda \approx$ 8.0 $\mu$m. {\bf e}, Temperature dependence of $R^{\rm NL}_n$ at 8 T for $B \parallel I \parallel a$-axis. } 
     \label{fig:1}
\end{figure}

The charge transport phenomena in solids usually obey the Ohm's law, which describes the local transport. When nonlocal transport is observed, it is often a signature of elaborate physics beyond the Ohm's law. There are plenty of examples of nonlocal transport stemming from exotic phenomena: Quantized edge transport due to quantum Hall effect \cite{Halperin1982}, nonlocal transport due to edge states in quantum spin Hall insulator \cite{Roth2009} or in quantum anomalous Hall insulator \cite{Lippertz2022}, negative nonlocal resistance due to crossed Andreev reflection \cite{Lee2017, Rosdahl2018, Uday2024}, nonlocal voltage due to chiral anomaly \cite{Parameswaran2014, CZhang2017}, nonlocal resistance due to spin or valley Hall effect \cite{Abanin2011, Gorbachev2014, Sui2015, Shimazaki2015, Jungwirth2012, Sinova2015}, electronic hydrodynamics \cite{Levitov2016, Torre2015, Gorbar2018}, etc. In our transport studies of the topological semimetal ZrTe$_5$, we observe a new type of nonlocal transport that differs from all the examples mentioned above. In particular, the decay length of the nonlocality measured in ZrTe$_5$ increases linearly with the width of the sample without saturation, which is very different from most nonlocal phenomena where the diffusion length of the relevant degree of freedom (such as spin or chirality) dictates the decay length of the nonlocality that becomes independent of the sample size at a large scale. 



The material studied here, ZrTe$_5$, is a layered van-der-Waals material and each layer (that form the $ac$ plane) consists of highly conducting ZrTe$_3$ chains (that run along the $a$-axis) connected via additional Te atoms, leading to a strong transport anisotropy \cite{Weng2014, Lv2017, Wang2022}. Its low-energy physics is well described by a three-dimensional (3D) massive Dirac equation with a very small mass term \cite{Weng2014, RYChen2015, Li2016, Liu2016, YZhang2017, HWang2018, Liang2018, Li2018, Xu2018, Tang2019, Zhang2019, Sun2020, Fu2020, Galeski2021, Wang2022, Wang2023, Wang2024}, making it topological. Peculiarly, the chemical potential $\mu$ shifts with temperature, leading to a peak in the temperature dependence of the resistivity at $T_p$ where $\mu$ reaches the Dirac point \cite{Xu2018, YZhang2017, Fu2020}. When ZrTe$_5$ crystals are grown by a Te-flux method, one can obtain samples with $T_p \approx$ 0 K having a very low carrier density of $\sim$10$^{16}$ cm$^{-3}$ \cite{Wang2025, Wang2023, Wang2022, Liang2018, Sun2020, HWang2018}. Various novel transport properties have been reported for ZrTe$_5$, including chiral magnetic effect \cite{Li2016}, 3D quantum Hall effect \cite{Tang2019, Galeski2021}, unconventional anomalous Hall effect \cite{Liang2018, Sun2020}, gigantic magnetochiral anisotropy (MCA) \cite{Wang2022, Wang2024b}, peculiar nonlinear transport \cite{Wang2023}, and unusual parallel-field Hall effect \cite{Wang2025}; we note that the latter four phenomena point to the breaking of inversion symmetry and are observed in $T_p \approx$ 0 K samples. The samples used in this study are Te-flux-grown crystals showing $T_p \approx$ 0 K (see Ref. \cite{SM}). We studied both bulk single crystals and micron-sized flakes to explore a wide range of length scale. Note that in ZrTe$_5$ the usual $xyz$ transport axes correspond to $acb$ crystal axes (see Fig. 1a).

In this study, we primarily used the geometry shown in Fig. 1a to address the nonlocal transport. The current is injected from electrode 1 and drained through grounded electrode 2, setting the area beyond electrode 2 to be in the ``nonlocal" region; the nonlocal resistance is defined as $R^{\rm NL}_n \equiv (V_n - V_8) / I_{12}$, where $n = 3-6$ labels the electrodes and $I_{12}$ is the applied current. We observed a sizable $R^{\rm NL}_n$ when a large magnetic field $B$ was applied along the $a$-axis, and interestingly, a small out-of-plane misalignment of only $\sim$1$^{\circ}$ makes $R^{\rm NL}_n$ to disappear (Fig. 1b). The negative sign of $R^{\rm NL}_n$ means that the potential at electrode $n$ is lower than that at electrode $8$. The $B$-dependence of $R^{\rm NL}_n$ is predominantly symmetric and exhibits an onset threshold of $\sim$2 T, which roughly corresponds to the onset of ultraquantum limit \cite{Wang2022, Wang2023}. Importantly, $R^{\rm NL}_n$ decays exponentially with $x$ in the nonlocal region as shown in Fig. 1d, where $x$ is the distance measured from the grounded electrode 2 to the nonlocal voltage electrode $n$ (= $3-6$). The decay length $\lambda$ obtained from the fit in Fig. 1d is 8.0 $\mu$m, which is surprisingly long. 
These data were obtained on a flake-based device sample D1 shown in Fig. 2b whose thickness was 0.38 $\mu$m (see Ref. \cite{SM} for fabrication details). 
The temperature dependence of $R^{\rm NL}_n$ is non-monotonic and it persists up to $\sim$100 K (Fig. 1e), which is different from the MCA or the nonlinear transport that only show up at low temperature \cite{Wang2022, Wang2023}. The data in Figs. 1b-1d were taken at 48 K, where $|R^{\rm NL}_n|$ was maximum.


\begin{figure}[t]
	\centering
	\includegraphics[width=8.6cm]{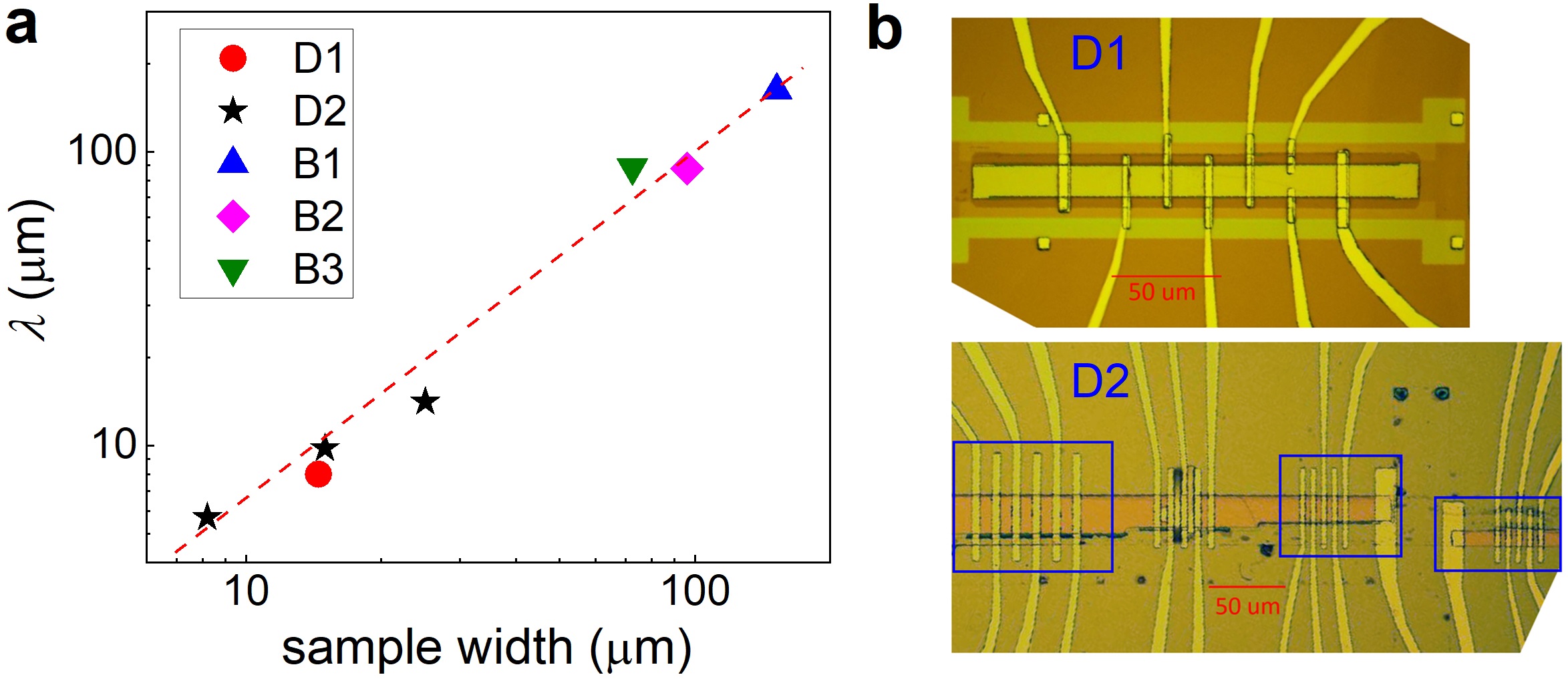}
	\caption{{\bf Sample-width dependence of the decay length.} {\bf a}, Decay length of the nonlocal resistance measured on various samples in 8 T for $B \parallel I \parallel a$-axis at the temperature where $|R^{\rm NL}_n|$ was maximum (see Ref. \cite{SM} for the data and analyses) as a function of the sample width. Samples D1 and D2 are flake-based devices, while samples B1, B2 and B3 are bulk single crystals. The red dashed line represents a linear relationship in this double-logarithmic plot. {\bf b}, Optical images of samples D1 and D2. In the fabrication of sample D2, a flake was etched to create three sections (marked by blue rectangles) with the same thickness but varying widths.} 
     \label{fig:2}
\end{figure}

 To gain more insight into this unusual nonlocal transport, we measured many samples with different dimensions, and we discovered a strong correlation between $\lambda$ and the width $w$ of the sample:  Essentially, $\lambda$ increases linearly with $w$ as shown in Fig. 2a over more than a decade (see Ref. \cite{SM} for supporting data). The samples B1--B3 with a large $w$ were bulk single crystals cleaved into thin rectangles. The device sample D2 had three sections that were etched into different widths but with the same thickness (see Fig. 2b). The observed variation in $\lambda$ for the three sections of sample D2 indicates that the thickness is not the decisive factor for $\lambda$ but the width is. The sample-width dependence of $\lambda$ implies that the system is not translationally invariant along the $y$ direction. For example, if opposite charges accumulate on the opposing edges in the nonlocal region (which could happen if a current component along $z$ is generated in the nonlocal region and creates a Hall voltage due to $B \parallel x$), the resulting electric field breaks translational invariance along $y$; such an electric field would be detected as a ``nonlocal Hall effect".


To test this possibility, we changed the measurement configuration to that shown in Fig. 3a such that the electrical current $I_{1n}$ injected from electrode 1 was drained at electrode $n$ ($=2-5$), while the nonlocal transverse voltage was always measured between electrodes 6 and 7. The nonlocal Hall resistance, defined as $R^{\rm NL}_{{\rm Hall},n} \equiv (V_6 - V_7) / I_{1n}$, was indeed observed in sample D1 for $B \parallel I \parallel a$-axis with a nontrivial $B$-dependence (Fig. 3b). We took the temperature dependence of $R^{\rm NL}_{{\rm Hall},n}$ at 2 T (Fig. 3c) where its size is the largest; this temperature dependence is very similar to that of $R^{\rm NL}_n$ shown in Fig. 1e, pointing to a common origin. The distance dependence (Fig. 3d) is also the same as that for the longitudinal signal. 

\begin{figure}[t]
	\centering
	\includegraphics[width=8.5cm]{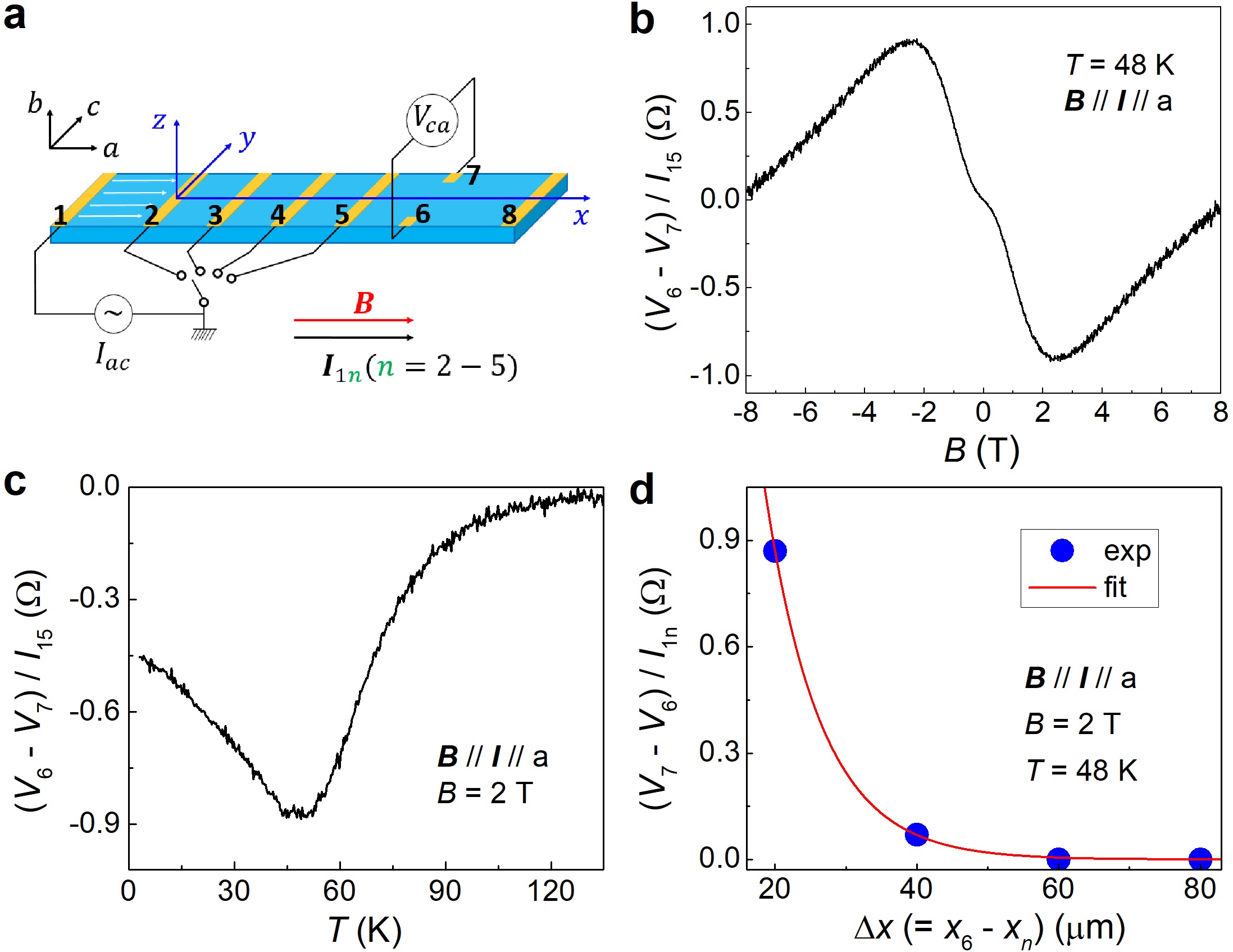}
	\caption{{\bf Nonlocal Hall effect in sample D1.} {\bf a}, Schematic drawing of the measurement configuration. Current $I_{1n}$ was injected from electrode 1 and drained through the grounded electrodes $n$ (= $2-5$). The magnetic field $B$ was applied along the $a$-axis. {\bf b}, $B$-dependence of the nonlocal Hall resistance for $n$ = 5, $R^{\rm NL}_{{\rm Hall},5} = (V_6 - V_7)/I_{15}$, at 48 K. The $R^{\rm NL}_{{\rm Hall},n}$ data were antisymmetrized to remove the admixture of the longitudinal resistance due to contact misalignment. {\bf c}, Temperature dependence of $R^{\rm NL}_{{\rm Hall},5}$ for $B \parallel I \parallel a$-axis at 2 T. {\bf d}, Dependence of $R^{\rm NL}_{{\rm Hall},n}$ ($n = 2-5$) on $\Delta x$ (= $x_6 - x_n$) at 2 T and 48 K for $B \parallel I \parallel a$-axis; $x_n$ is the coordinate position of the grounded electrode $n$. Red line is a fit to the formula $R^{\rm NL}_{{\rm Hall},n} = R_0 e^{-x/\lambda}$, which gives $\lambda \approx$ 7.9 $\mu$m.}
	\label{fig:3}
\end{figure}

We note that, although such a nonlocal Hall effect can in principle happen due to some inhomogeneous current distribution, our observation cannot be understood in this way. This is highlighted by the observation of a sizable $R^{\rm NL}_{{\rm Hall},n}$ signal for $n$ = 4, which means that a sizable electric field is generated between electrodes 6 and 7 as a result of the local current between electrodes 1 and 4, despite the existence of the electric short in-between caused by electrode 5. This unusual nonlocal Hall effect was also reproduced in another sample (see Ref. \cite{SM}).

Next we present how $R^{\rm NL}_n$ changes when $B$ is rotated away from the $a$-axis in the $ac$ plane (i.e. $xy$ plane) by an angle $\varphi$. Since $R^{\rm NL}_n$ is very sensitive to the out-of-plane misalignment of $B$, for this measurement we first fixed $\varphi$ and swept $\theta$ in a narrow range in the vertical $xz$ plane across the $xy$ plane, such that we can identify the maximum in $|R^{\rm NL}_n|$ that occurs when $\theta$ is exactly 90$^{\circ}$. More details about this type of measurements are found in Ref. \cite{Wang2025}. The observed $B$-dependences of $R^{\rm NL}_n$ for various $\varphi$ values are unusual in that they develop an antisymmetry as $\varphi$ gets closer to 90$^{\circ}$ (Fig. 4a). On the other hand, $R^{\rm NL}_{{\rm Hall},n}$ was found to weaken as $\varphi$ approaches 90$^{\circ}$ (Fig. 4b). The development of asymmetry in $R^{\rm NL}_n(B)$ as $B$ approaches the $y$-axis suggests the involvement of magnetic moments in this direction.

\begin{figure}[t]
	\centering
	\includegraphics[width=8.6cm]{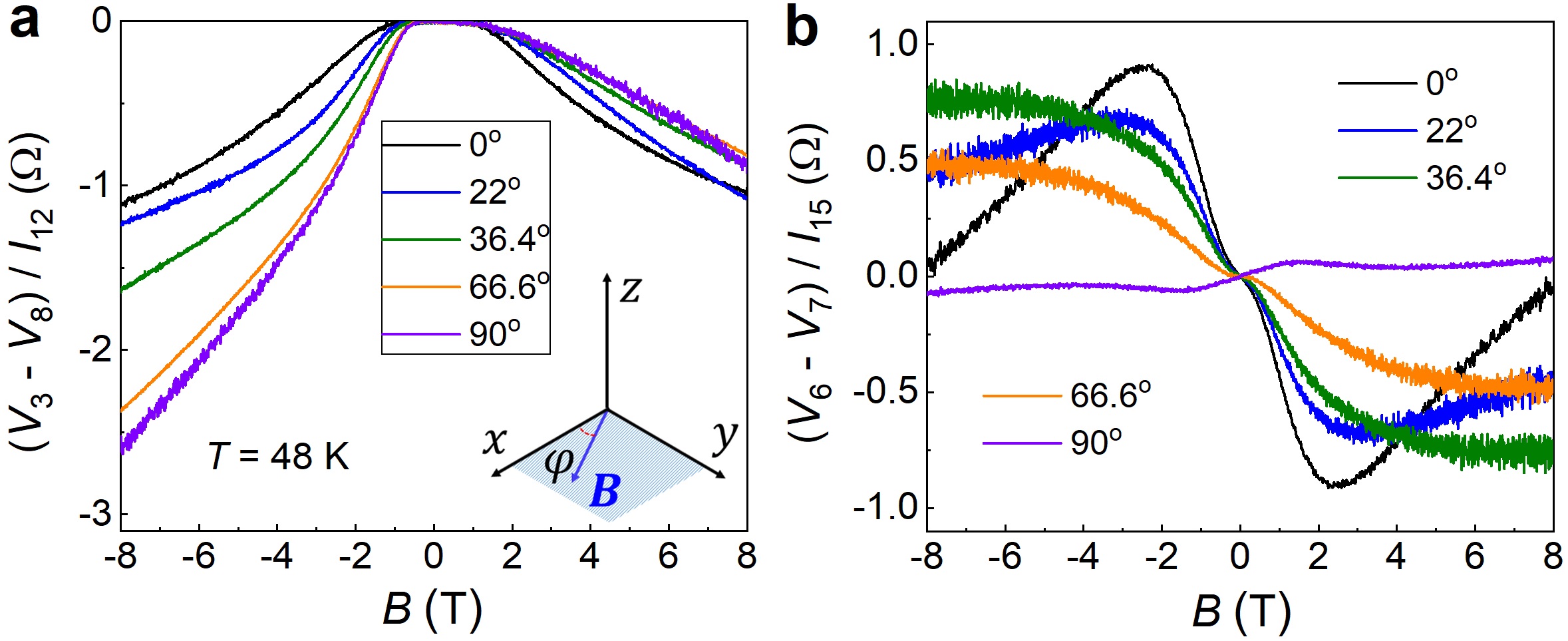}
	\caption{{\bf Nonlocal transport in sample D1 for $B$ rotated in the $xy$ plane.} {\bf a}, $B$-dependences of $R^{\rm NL}_3$ [= $(V_3 - V_8) / I_{12}$] for the configuration of Fig. 1a at 48 K for various magnetic-field angles $\varphi$ varied in the $ac$ plane. Inset shows the definition of $\varphi$. {\bf b}, $B$-dependences of $R^{\rm NL}_{{\rm Hall},5}$ for the configuration of Fig. 3a at various $\varphi$ at 48 K. }
	\label{fig:4}
\end{figure}



To put our result in context, we have performed additional experiments in the standard nonlocal transport configuration used for the studies of the chiral currents in a Weyl semimetal \cite{CZhang2017} or the spin/valley currents in graphene \cite{Abanin2011, Gorbachev2014, Sui2015, Shimazaki2015}. We found that the configuration with $I \parallel c$ (sample B4) yielded inconclusive results due to the non-negligible Ohm's-law contribution amplified by the higher conductivity along $a$ (the details are described in Ref. \cite{SM}). On the other hand, the result for the standard configuration with $I \parallel b$ (sample B5) was largely consistent with the result for $I \parallel a$ and showed that the existence of $B \parallel a$ is crucial for this nonlocal effect; this means that $B \parallel I$ is {\it not} a necessary condition, but $B \parallel a$ is. In addition, we have performed an experiment on a ZrTe$_5$ sample with $T_p \approx$ 108 K and found that the unusual nonlocal transport is absent in such a sample (see Ref. \cite{SM}). This suggests that the tuning of the chemical potential close to the Dirac point is necessary for this phenomenon to take place.


One may wonder if the observed nonlocal voltage might be explained by the Ohm's law as a result of inhomogeneous current distribution induced by the current electrodes that are made on the top surface; namely, due to the finite thickness, the electric current naturally diffuses into the nonlocal region, leading to a voltage drop in the nonlocal region. The large resistivity anisotropy would enhance its plausibility. To examine this scenario, we performed 3D COMSOL simulation based on Ohm's law to calculate the expected current and voltage distributions (see Ref. \cite{SM}). This simulation gives a very short decay length of less than 1 $\mu$m in the nonlocal region for the parameters of sample D1. Moreover, the simulation cannot reproduce the almost singular magnetic-field-orientation dependence as well. Therefore, one can safely rule out the inhomogenous current distribution obeying the Ohm's law as the cause of the observed nonlocal transport.
 Our 3D COMSOL simulation also rules out the possibility of the trivial origin of the observed non-local Hall effect (see Ref. \cite{SM}). 

Another trivial cause of the nonlocal voltage is a thermal effect. Local Joule heating creates a temperature gradient in the nonlocal region, which induces a voltage through Seebeck effect that would present an exponential decay. However, the nonlocal voltage caused by Joule heating should increase quadratically with current, but we observed a linear dependence (see Ref. \cite{SM}), which clearly rules out the thermal origin of the observed nonlocal voltage. To firm up this conclusion, we performed a thermal simulation and obtained the decay length of $\sim$ 1.4 $\mu$m for the Joule-heating-induced nonlocal voltage in sample D1 (see Ref. \cite{SM}), which is much shorter than the experiment. 
Furthermore, our thermal simulation shows that the possible temperature rise due to Joule heating in sample D1 is only less than 0.01 mK in the experimental situation (see Ref. \cite{SM}). Therefore, trivial origins of nonlocal voltage cannot explain our observation.

A nonlocal transport in the standard configuration with a long decay length $L_{\rm v}$ (a few $\mu$m or longer) was predicted for Weyl semimetals due to chiral anomaly \cite{Parameswaran2014}, and an experiment performed on Cd$_3$As$_2$ found $L_{\rm v} \approx$ 1.5 $\mu$m \cite{CZhang2017}. For this nonlocal transport, the existence of a pair of Weyl cones having opposite chiralities is a prerequisite. However, in ZrTe$_5$ no Weyl cones are generated with $B \parallel a$ \cite{Wang2022}, and hence this mechanism is not likely to be relevant in our case. Also, even though the Dirac cone in ZrTe$_5$ may split into two Weyl cones with $B \parallel c$, our measurement of sample B4 mentioned above for the standard nonlocal configuration found a minimum in the nonlocal signal for $B \parallel I \parallel c$ (see Ref. \cite{SM}). 
Another possible mechanism for nonlocal transport is the spin Hall effect \cite{Sinova2015}, but in ZrTe$_5$ the spin-orbit coupling is strong and the spin diffusion length should be of the order of the mean-free path, which is $\sim$40 nm \cite{Wang2022} and is much shorter than $\lambda$.
Finally, edge states provide a mechanism for nonlocal transport, but our observations of the exponential decay of the nonlocality and the width-dependence of the decay length both speak against the edge-state scenario. Therefore, none of the known mechanisms of nonlocal transport can explain our result. 

Although the origin of this phenomenon remains elusive, our experimental observations provide a couple of important clues for understanding its underlying mechanism. 
First, the exponential decay length $\lambda$ scales linearly with the sample width $w$, suggesting a nonlocal voltage of the form $V^{\rm NL} \sim e^{-\beta x/w}$, which is reminiscent of two-dimensional diffusion between a source and a sink located at the edges of a strip (see Ref. \cite{SM}). The emergence of a voltage difference in the nonlocal region may involve a conversion mechanism akin to the inverse spin Hall effect \cite{Sinova2015}, similar to the case of chirality diffusion \cite{Parameswaran2014, CZhang2017}.

Second, the exceptionally large decay length $\lambda$ (exceeding 100 $\mu$m) suggests an extremely low scattering rate for the involved transport channel. This condition might be expected in the ultraquantum regime, where carriers are confined to the lowest Landau level, the torus-shaped Fermi surface of ZrTe$_5$ collapses into quasi-1D flat bands \cite{Wang2023}, and the electronic structure becomes effectively one-dimensional (1D) along the magnetic field. Such dimension reduction, combined with the intrinsic anisotropy of ZrTe$_5$, may drastically suppress transverse scattering processes in this measurement configuration. 



In summary, we discovered unusual nonlocal transport response in the nodal semimetal ZrTe$_5$ in the configuration $B \parallel I \parallel a$ for magnetic fields that bring the system into the ultra-quantum limit. The nonlocal voltage presents an exponential decay, and the decay length scales linearly with the sample width without saturation, suggesting vanishingly small scattering for the mechanism that lies behind the nonlocality. An unusual nonlocal Hall voltage, that can hardly be of trivial origin, was also observed. Whatever the mechanism behind this striking phenomenon is, the present experiment demonstrates that nonlocal transport is a useful tool in the study of topological quantum materials.

\section{acknowledgements}
We thank A. Rosch, T. B\"omerich, J. Park, and H. F. Legg for helpful discussions. This work has received funding from the Deutsche Forschungsgemeinschaft (DFG, German Research Foundation) under CRC 1238-277146847 (subprojects A04 and B01) and also from the DFG under Germany’s Excellence Strategy -- Cluster of Excellence Matter and Light for Quantum Computing (ML4Q) EXC 2004/1-390534769. 

{\bf Data availability:} The data that support the findings of this study are available at the online depository zenodo with the identifier {10.5281/zenodo.15330418}. 



\bibliography{ZrTe5_bibliography}

\end{document}




\title{Supplementary Materials for 
``Magnetic-field-induced nonlocal transport in the topological semimetal ZrTe$_5$''}

\author{Yongjian Wang}
\affiliation{Physics Institute II, University of Cologne, D-50937 K\"oln, Germany}

\author{A. A. Taskin}
\affiliation{Physics Institute II, University of Cologne, D-50937 K\"oln, Germany}

\author{Yoichi Ando}
\affiliation{Physics Institute II, University of Cologne, D-50937 K\"oln, Germany}

\maketitle

\section{Materials and Methods}

Single crystals of ZrTe$_5$ with $T_p \approx$ 0 K were grown using a Te flux method. The samples with $T_p$ = 108 K were grown by a chemical vapor transport method. The details were described in Ref. \cite{Wang2022}. For bulk samples used for the transport measurements, the surface was cleaned by Ar plasma to remove the oxidized layer, and Au/Nb contact electrodes were sputter-deposited, after which gold wires were bonded to the sample with silver paste. For the device samples, the ZrTe$_5$ flakes were mechanically exfoliated from high-quality single crystals and transferred to a wafer (SiO$_2$ 290 nm on doped Si). The fabrication process of the devices was reported in Ref. \cite{Wang2025}. In fact, the sample D1 was also used in Ref. \cite{Wang2025}. Transport measurements were performed in a Quantum Design Physical Properties Measurement System (PPMS) with a rotating sample holder. The transport data were measured using a low-frequency (17.33 Hz) AC lock-in technique. The DC data ($I$-$V$ curves) for sample B1 were measured with Keithley 2450 source meter as DC current source and Keithley 2182 as voltage meter. The simulations for the voltage and temperature distributions were performed using the \textit{Electric Currents} interface within the \textit{AC/DC Module} and the \textit{Heat Transfer in Solids} interface within the \textit{Heat Transfer Module} of \textsc{COMSOL Multiphysics\textsuperscript{\textregistered}} (version~5.3), respectively.

\section{Basic characterization of sample D1}

\begin{figure}[h]
	\centering
	\includegraphics[width=9 cm]{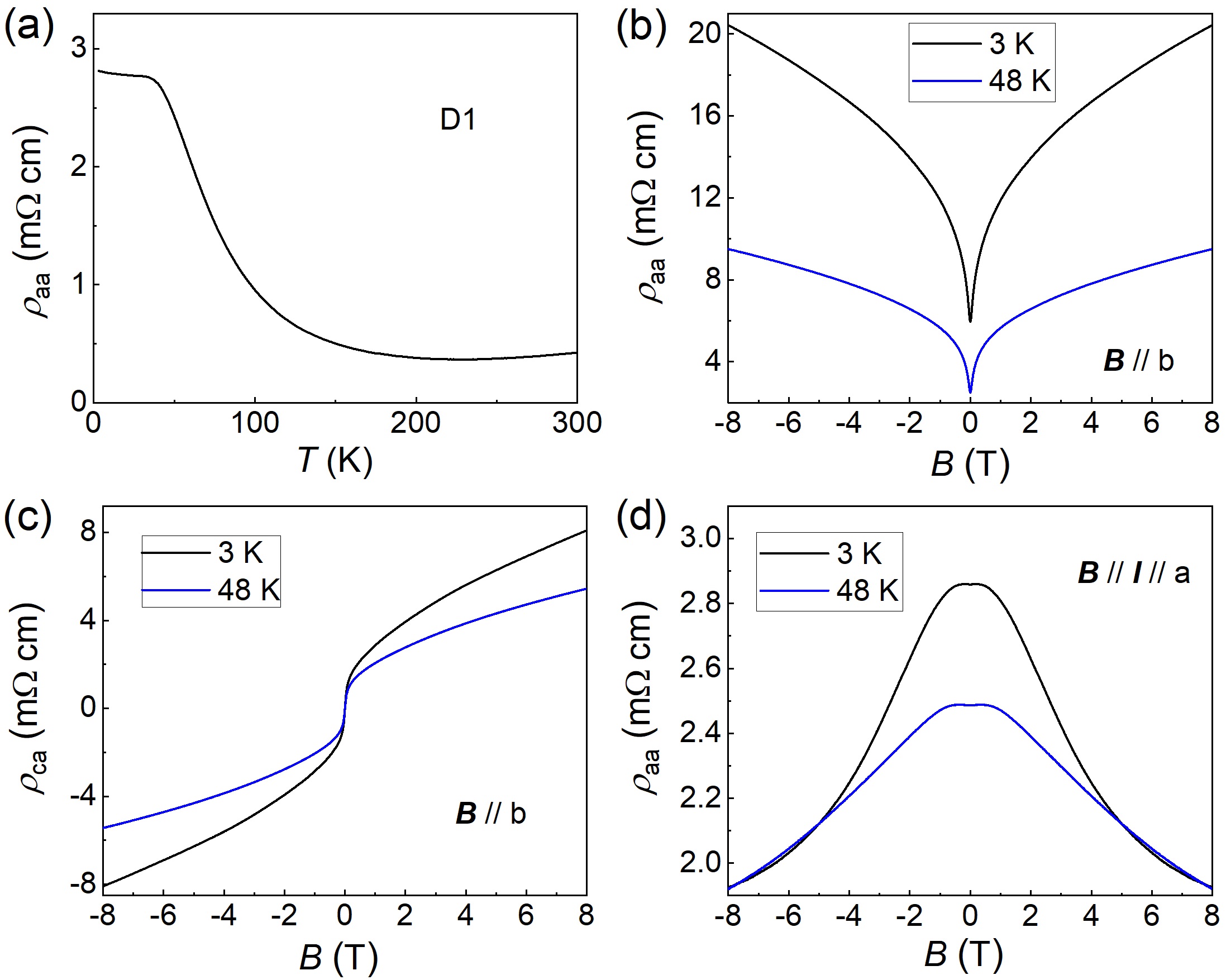}
	\caption{\textbf{Basic transport properties of sample D1.} (a) Temperature dependence of $\rho_{aa}$. (b, c) $B$-dependence of $\rho_{aa}$ (b) and $\rho_{ca}$ (c) at 3 K and 48 K for $B \parallel b$-axis. (d) $B$-dependences of $\rho_{aa}$ at 3 K and 48 K for $B \parallel I \parallel a$-axis.  
	}
	\label{fig:basic transport}
\end{figure}

The temperature dependence of the resistivity in ZrTe$_5$ generally shows a peak at a characteristic temperature $T_p$, marking a sign reversal of the charge carriers and Lifshitz transition \cite{Xu2018, YZhang2017}. The samples with $T_p \approx$ 0 K exhibit interesting physical properties at low temperature, such as magnetochiral anisotropy and nonlinear transport, due to the nodal-line semimetal nature realized in such samples \cite{Wang2022, Wang2023}. When making a device based on ZrTe$_5$, the $T_p$ is very sensitive to fabrication processes. By optimizing the device fabrication process, we successfully achieved ZrTe$_5$ devices with $T_p \approx$ 0 K [Fig. \ref{fig:basic transport}(a)]. The peculiar magnetoresistance and anomalous Hall effect observed in single crystal samples \cite{Wang2022} were reproduced in our device [Fig. \ref{fig:basic transport}(b) and (c)]. The observation of parallel-field Hall effect for $B \parallel I \parallel a$-axis reported in Ref. \cite{Wang2025} was actually performed in sample D1 of the present study. 
In the samples with $T_p \approx$ 0 K, the small quantum-limiting field ($\sim$ 0.5 T) and large anomalous Hall effect make it challenging to extract the carrier density $n$ from $\rho_{ca}(B)$ \cite{Liang2018, Wang2022}. Hence, $n \simeq 10^{16}$ cm$^{-3}$ was estimated from quantum oscillations \cite{Wang2022}.



\section{Decay length $\lambda$ of the nonlocal resistance in various samples}

\begin{figure}[h]
	\centering
	\includegraphics[width=15 cm]{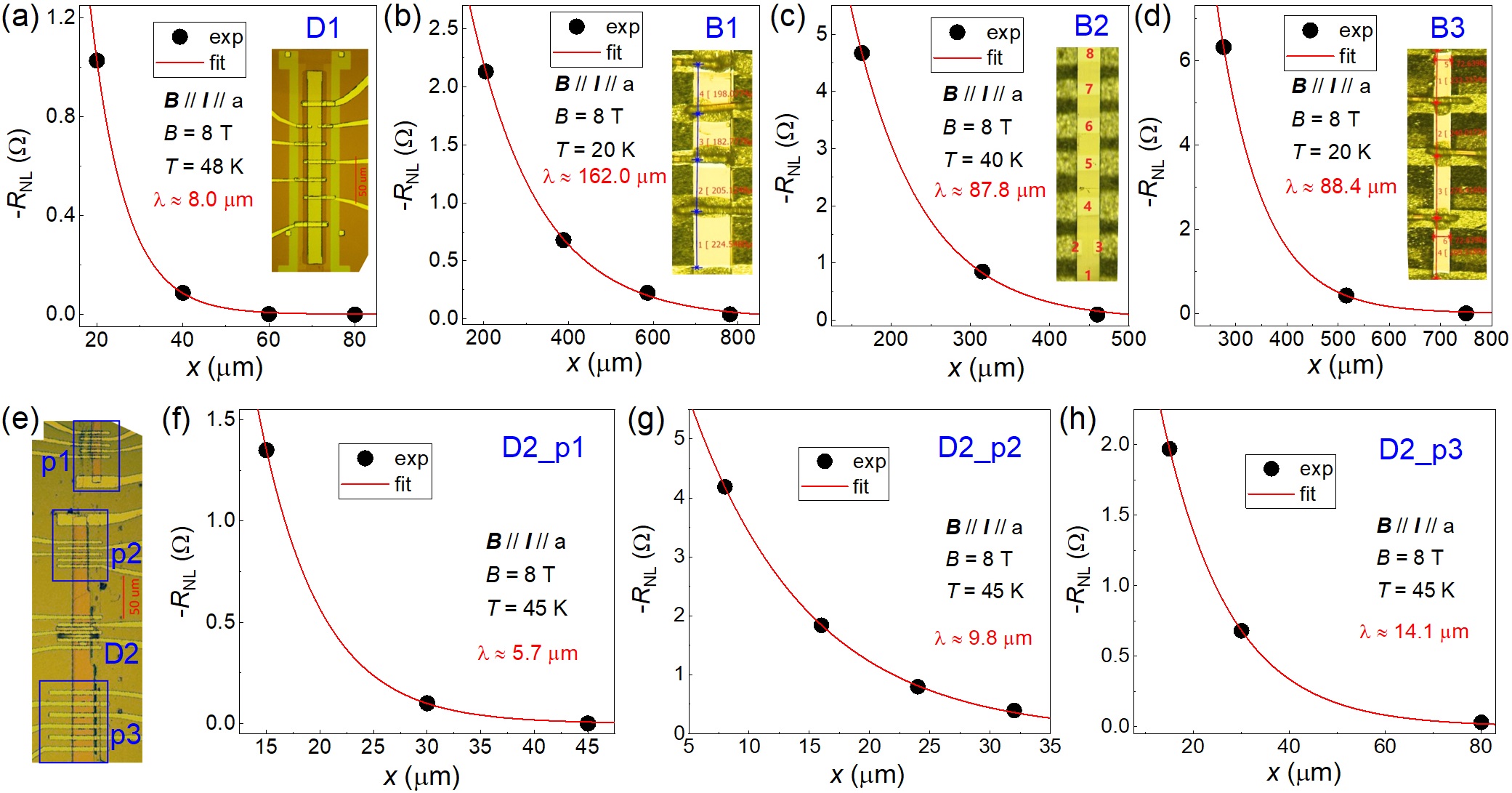}
	\caption{\textbf{Decay length of the nonlocal resistance measured in various samples.} (a-d) $x$-dependence of $-R_{\rm NL}$ at 8 T for $B \parallel I \parallel a$-axis, where $x$ is the distance measured from the grounded electrode to the nonlocal voltage electrode. The insets are the optical images of corresponding samples. (e) Optical image of sample D2, where a flake was etched to form sections (labeled p1--p3) with the same thickness but varying widths, as marked by blue rectangles. (f-h) $x$-dependence of $-R_{\rm NL}$ at 8 T for $B \parallel I \parallel a$-axis for different portions of sample D2 marked in (e).	
	}
	\label{fig:various samples}
\end{figure}

The nonlocal transport was reproduced in all the samples with $T_p \approx$ 0 K, including fabricated devices (D1 and D2) and bulk single crystals (B1--B3), and the decay length $\lambda$ was extracted for all the samples [Fig. \ref{fig:various samples}]. Table~\ref{table:1} summarizes the sample dimensions and $\lambda$ of the samples presented in Fig. \ref{fig:various samples}. While $\lambda$ shows a linear dependence on the width $w$ as shown in the main text (Fig. 2a), there is no correlation between $\lambda$ and the thickness $d$.

\setlength{\tabcolsep}{3mm}
\begin{table}[h]
		\caption{Summary of the samples in Fig. \ref{fig:various samples}. The parameters are: width $w$, thickness $d$, decay length $\lambda$.}
		\label{table:1}
		{\small
			\begin{center}
				\begin{tabular}{c|ccccccc}
				 Device    & D1   & D2-p1 & D2-p2 & D2-p3 & B1     & B2    & B3    \\ \hline
				 $w$ ($\mu$m)    & 14.5  & 8.2   & 15.0  & 25.1  & 152.2  & 96.0  & 72.6  \\
				 $d$ ($\mu$m)    & 0.38  & 0.36  & 0.36  & 0.36  & 8.2    & 18.1  & 4.9   \\
				 $\lambda$ ($\mu$m) & 8.0  & 5.7   & 9.8   & 14.1  & 162.0  & 87.8  & 88.4 \\
				\end{tabular}
			\end{center}
		}
	\end{table}


\section{Raw data of the nonlocal Hall effect in sample D1}

\begin{figure}[h]
	\centering
	\includegraphics[width=11 cm]{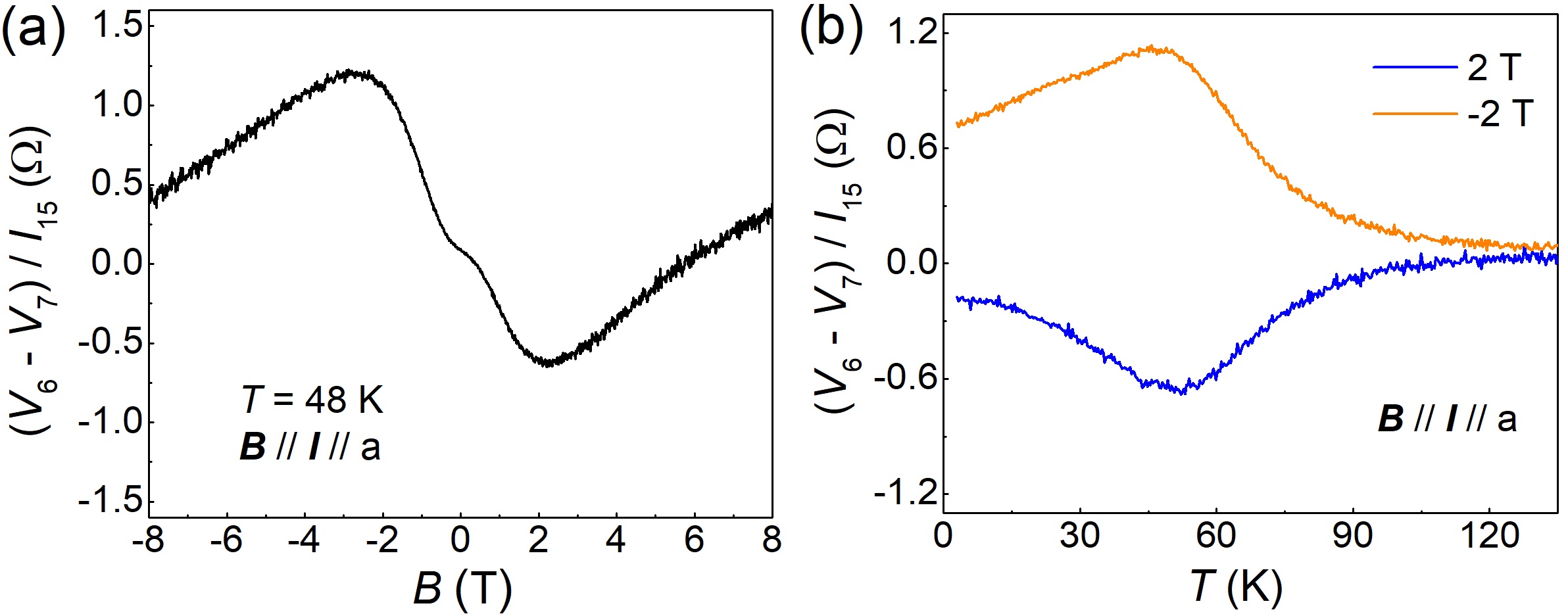}
	\caption{\textbf{Raw data of the nonlocal Hall effect in sample D1.} (a) $B$-dependence of $(V_6-V_7)/I_{15}$ at 48 K for $B \parallel I \parallel a$ without antisymmetrization. (b) Temperature dependence of $(V_6-V_7)/I_{15}$ in $\pm$2 T of $B \parallel I \parallel a$ without antisymmetrization.
	}
	\label{fig:raw data of the nolocal Hall in D1}
\end{figure}

Here we present the raw data of the nonlocal Hall effect observed in sample D1 with the current $I_{15}$ before antisymmetrization. As one can see in Fig. \ref{fig:raw data of the nolocal Hall in D1}, the raw data are already dominated by the antisymmetric component. The nonlocal Hall data presented in the main text have been antisymmetrized to remove a small admixture of the longitudinal resistance due to contact misalignment.

\section{Simulations}
\label{Sec:3D simulation}

\subsection{3D simulation of the inhomogeneous current distribution obeying the Ohm's law} 

Considering the finite thickness of the samples and the placement of the current contacts on the top surface, the electric current naturally diffuses into the nonlocal region in a limited range, which leads to the nonlocal transport dictated by the current- and electric-field distribution obeying the Ohm's law. To clarify this contribution, we performed 3D simulation to calculate the electric potential distribution in sample D1 with COMSOL Multiphysics simulation software. This simulation is based on the Ohm's law. The charge conservation is described by $\nabla \cdot \mathbf{j} + \partial \rho_{\text{charge}}/\partial t = 0$. For steady state, the charge density $\rho_{\text{charge}}$ does not change with time $t$, which leads to $\nabla \cdot \mathbf{j} = 0$. With Ohm's law $j = \sigma E$ and $E = -\nabla \phi$, one obtains $\nabla (\sigma \nabla \phi)=0$, which describes the spatial distribution of the electric potential $\phi$ as
\begin{equation}
\begin{split}
\sigma_{xx} \frac{\partial^2 \phi}{\partial x^2} + \sigma_{xy} \frac{\partial^2 \phi}{\partial x \partial y} + \sigma_{xz} \frac{\partial^2 \phi}{\partial x \partial z} + \sigma_{yx} \frac{\partial^2 \phi}{\partial y \partial x} + \sigma_{yy} \frac{\partial^2 \phi}{\partial y^2} + \sigma_{yz} \frac{\partial^2 \phi}{\partial y \partial z} + \sigma_{zx} \frac{\partial^2 \phi}{\partial z \partial x} + \sigma_{zy} \frac{\partial^2 \phi}{\partial z \partial y} + \sigma_{zz} \frac{\partial^2 \phi}{\partial z^2} = 0 \,\,.
\label{A1}
\end{split}
\end{equation}
For sample D1 at 48 K in $B \parallel I \parallel a$-axis of 8 T, our previous study \cite{Wang2025} has elucidated the full resistivity tensor as
\begin{equation}
\rho(B=8 {\rm T})=\left[\begin{array}{lll}
\rho_{x x} & \rho_{x y} & \rho_{x z} \\
\rho_{y x} & \rho_{y y} & \rho_{y z} \\
\rho_{z x} & \rho_{z y} & \rho_{z z}
\end{array}\right]
\approx \left[\begin{array}{lll}
1.94 & 0.266 & 1.21 \\
-0.266 & 6.34 & -13.50 \\
1.21 & 13.50 & 63.34
\end{array}\right] \,\, (\rm m \Omega \, cm).
\label{A8}
\end{equation}

This resistivity tensor is converted into the conductivity tensor with $\rho \sigma = 1$ to obtain
\begin{equation}
\begin{split}
\sigma \approx \left[\begin{array}{lll}
52063.50 & -45.61 & -1005.53 \\
45.61 & 10854.91 & 2312.39 \\
-1005.53 & -2312.39 & 1105.33
\end{array}\right] \,\, (\rm \Omega^{-1} \, m^{-1}),
\label{A9}
\end{split}
\end{equation}
which is used as the inputs for the simulation of the electric potential distribution. 

\begin{figure}[h]
	\centering
	\includegraphics[width=10 cm]{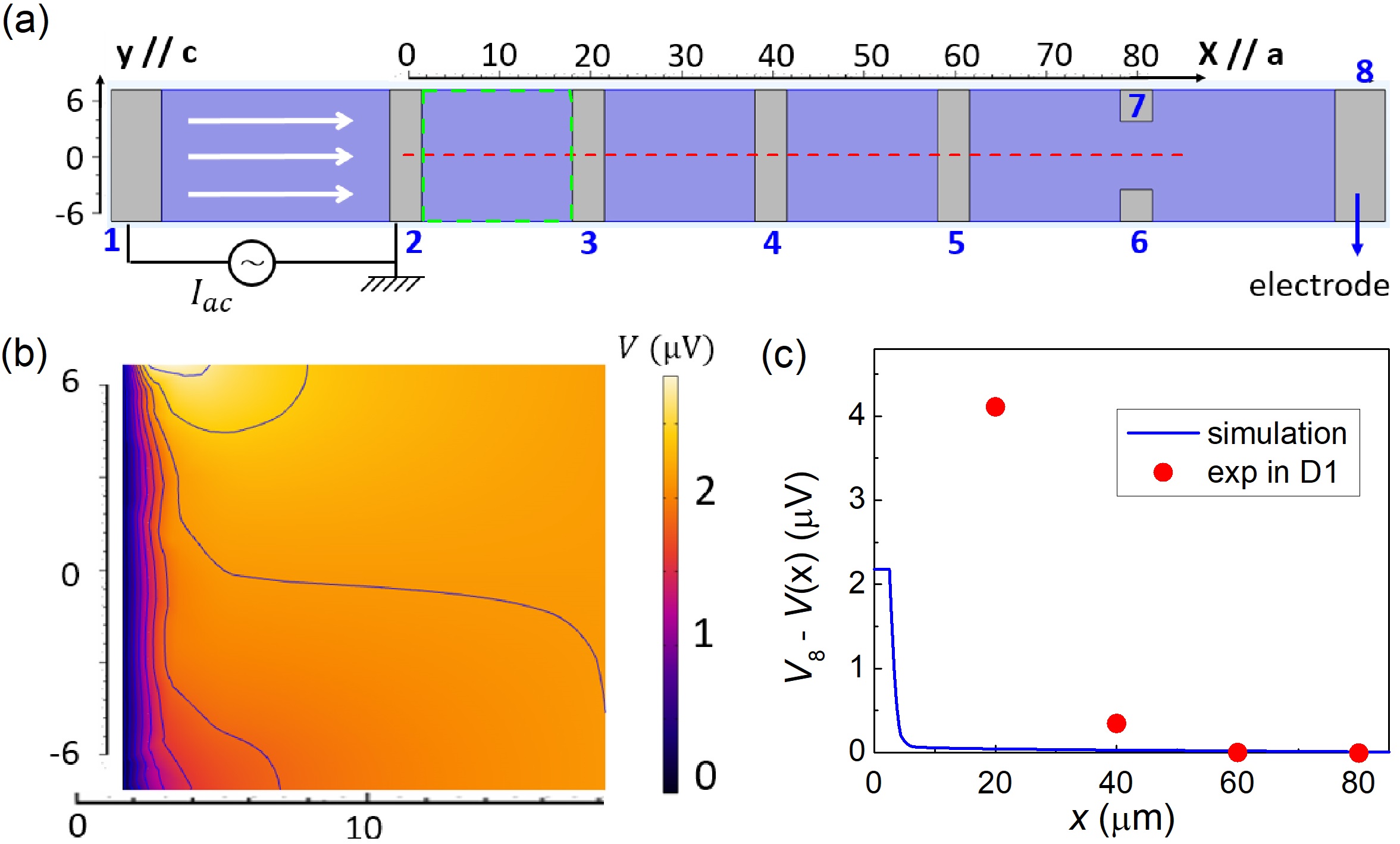}
	\caption{\textbf{3D Ohm's-law simulation for sample D1.} (a) Top view of the electrode configuration used in the simulation. The purple part is the sample and the gray strips are the electrodes. The electric current ($I$ = 4 $\mu$A) was injected from electrode 1 and drained through electrode 2. (b) Voltage distribution on the sample surface in the nonlocal region between electrodes 2 and 3 as marked by green-dashed box in (a). (c) $x$-dependence of the voltage difference $V_8 - V(x)$ on the surface along the red-dashed line shown in (a); the experimental data are shown here for comparison.
	}
	\label{fig:3D simulation}
\end{figure}

The 3D boundary condition for the simulation was set by the measurement configuration of sample D1, whose top view is shown in Fig. \ref{fig:3D simulation}(a). An electric current ($I$ = 4 $\mu$A) was injected from electrode 1 and drained through the grounded electrode 2. The magnetic-field effect is already incorporated in the conductivity tensor used for the simulation. The $xy$ coordinate system is defined on the top surface ($ac$ plane) of the sample. The result shows a fast decay of the voltage gradient in the nonlocal region [Figs. \ref{fig:3D simulation}(b) and \ref{fig:3D simulation}(c)] within 1 $\mu$m. 
Thus, the current- and electric-field distribution dictated by the Ohm's law cannot explain the unusual nonlocal transport with $\lambda \approx$ 8~$\mu$m observed in sample D1. 

Note that the weak asymmetry along the $y$ axis in Fig. \ref{fig:3D simulation}(b) is due to the finite $\rho_{yz}$ and $\rho_{yx}$ \cite{Wang2025} and it corresponds to the nonlocal Hall effect. The simulation of this effect is discussed in Sec. \ref{Sec:3D simulation}C.   


\subsection{Effects of thickness variation on the current distribution}  

\begin{figure}[h]
	\centering
	\includegraphics[width=10 cm]{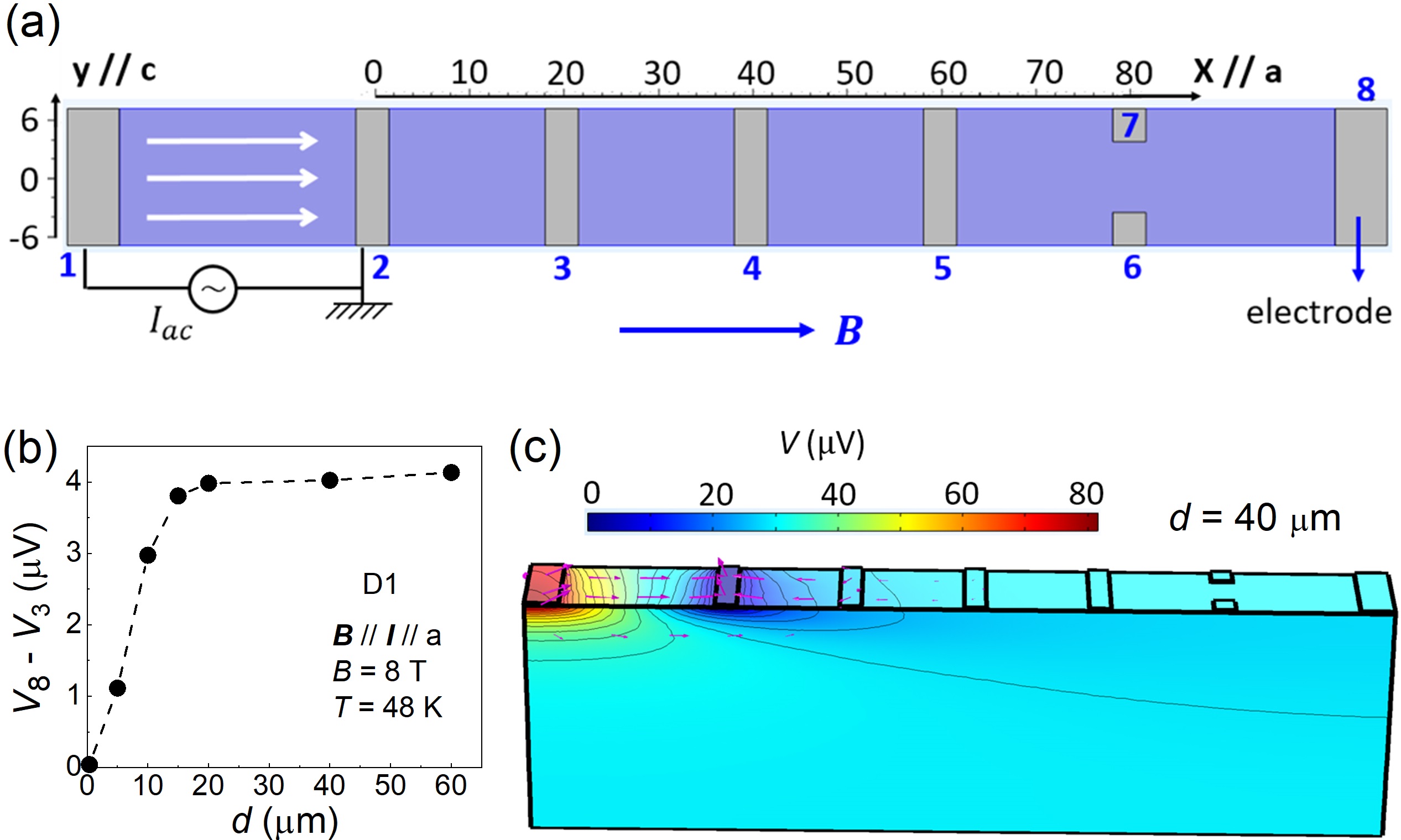}
	\caption{\textbf{3D Ohm's-law simulations of thicker samples.} (a) Top view of the electrode configuration used in the simulation. The purple part is the sample and the gray strips are the electrodes. The electric current ($I$ = 4 $\mu$A) was injected from electrode 1 and drained through electrode 2. (b) Thickness dependence of the voltage difference $V_8 - V_3$ between electrodes 8 and 3 obtained from simulations. (c) Current and voltage distributions in the case of $d$ = 40 $\mu$m. 
	}
	\label{fig:simulation on varying thickness}
\end{figure}

To investigate how the electric-current diffusion into the nonlocal region obeying the Ohm's law changes with thickness, simulations were performed for varying thickness $d$ while keeping other parameters [Fig. \ref{fig:simulation on varying thickness}]. As the thickness increases, the electric current obeying the Ohm's law diffuses more easily into the nonlocal region up to a threshold value of $\sim$20 $\mu$m, above which the current distribution (and hence the nonlocal voltage $V_8 - V_3$) hardly changes [Fig. \ref{fig:simulation on varying thickness}(b)]. The illustration of the situation for $d$ = 40 $\mu$m [\ref{fig:simulation on varying thickness}(c)] shows that the region of the sample deeper than $\sim$20 $\mu$m is free from current as well as voltage gradient. 

Note that the samples used in our experiments were very thin in comparison to the contact separations, which was the reason why the Ohm's-law contribution to the nonlocal voltage decays very rapidly. The thickness-dependence simulation suggests that the Ohm's-law contribution becomes important at the length scale of the current-contact separation when the thickness is also of the same order (or thicker).

\subsection{3D simulation of nonlocal Hall effect with $I_{14}$ in sample D1} 

\begin{figure}[h]
	\centering
	\includegraphics[width=10 cm]{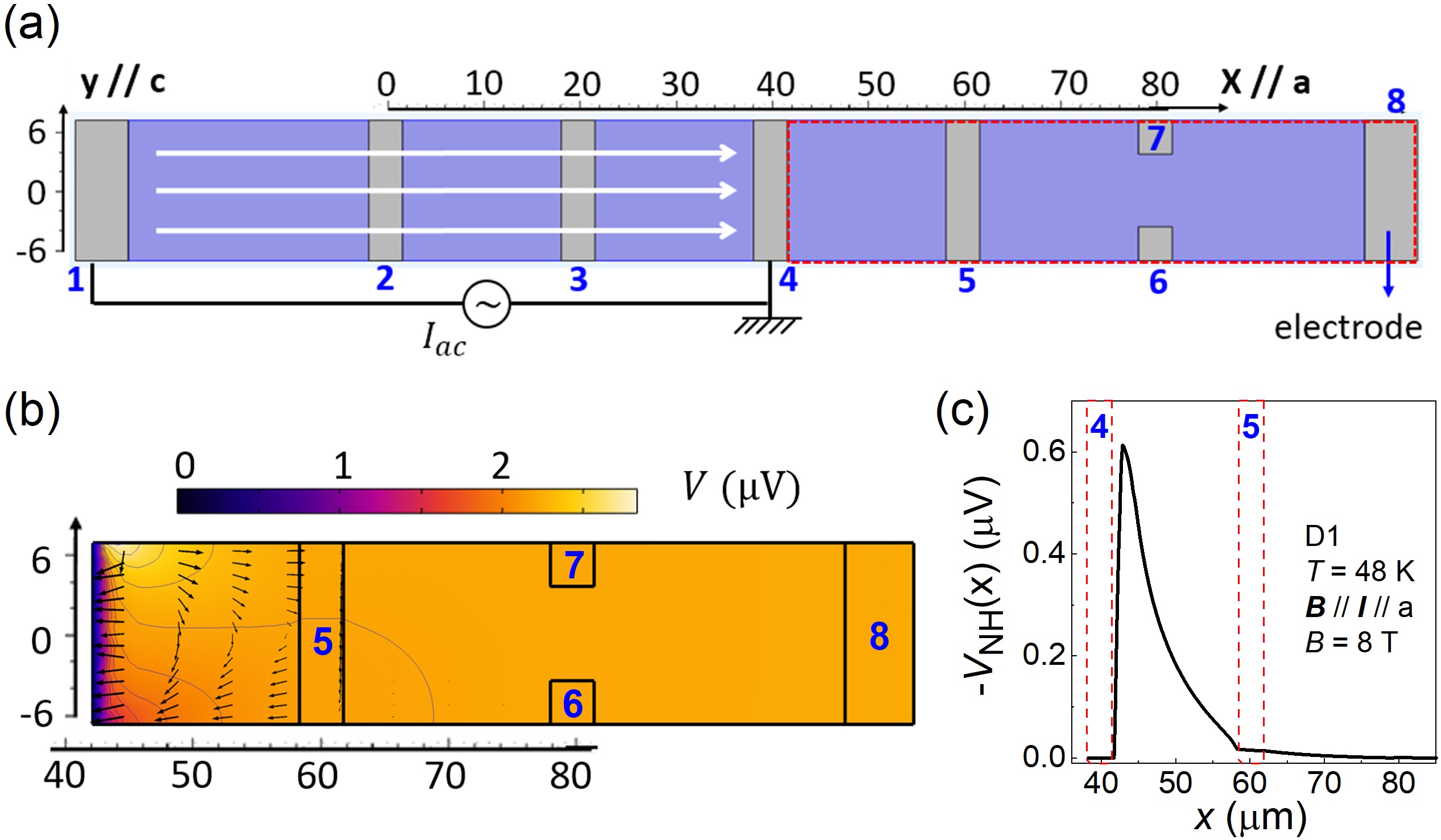}
	\caption{\textbf{3D Ohm's-law simulation of nonlocal Hall effect with $I_{14}$ in sample D1.} (a) Top view of the electrode configuration used in the simulation. The purple part is the sample and the gray strips are the electrodes. The electric current ($I$ = 4 $\mu$A) was injected from electrode 1 and drained through electrode 4. (b) Voltage distribution on the sample surface in the nonlocal region between electrodes 4 and 8 as marked by red-dashed box in (a). (c) $x$-dependence of the voltage difference $-V_{\rm NH}$ (voltage between the lower and upper edges). The position of the electrodes 4 and 5 are marked.
	}
	\label{fig:nonlocal Hall effect with shorting electrode}
\end{figure}

To examine the short-circuiting effect of the electrodes in the nonlocal Hall effect obeying the Ohm's law in sample D1, simulations were performed with an electric current ($I$ = 4 $\mu$A) injected from electrode 1 and drained through the grounded electrode 4 [Figs. \ref{fig:nonlocal Hall effect with shorting electrode}(a)]. Other than the choice of the current electrodes, the simulation configuration remains the same as that described in Section \ref{Sec:3D simulation}A. Due to the the finite $\rho_{yz}$ and $\rho_{yx}$ \cite{Wang2025}, the voltage difference $-V_{\rm NH}$ (voltage between the lower and upper edges) is created [Figs. \ref{fig:nonlocal Hall effect with shorting electrode}(b)]. However, this voltage difference is short-circuited at the elctrode 5, and the area beyond electrode 5 has virtually no transverse voltage gradient [Figs. \ref{fig:nonlocal Hall effect with shorting electrode}(c)]. As a result, the nonlocal Hall effect detected with electrodes 6 and 7 (corresponding to $x$ = 80~$\mu$m) is virtually zero ($\sim$ 4 $\times 10^{-4}$ $\mu$V) in this simulation, whereas we observed a sizable $V_6 - V_7$ of 0.28 $\mu$V in our experiment. This simulation confirms that when a shorting electrode is present, the Ohm's-law contribution to the nonlocal Hall effect vanishes. 

\subsection{Spurious nonlocal transport induced by Joule heating} \label{Sec:Joule heating}

\begin{figure}[h]
	\centering
	\includegraphics[width=10 cm]{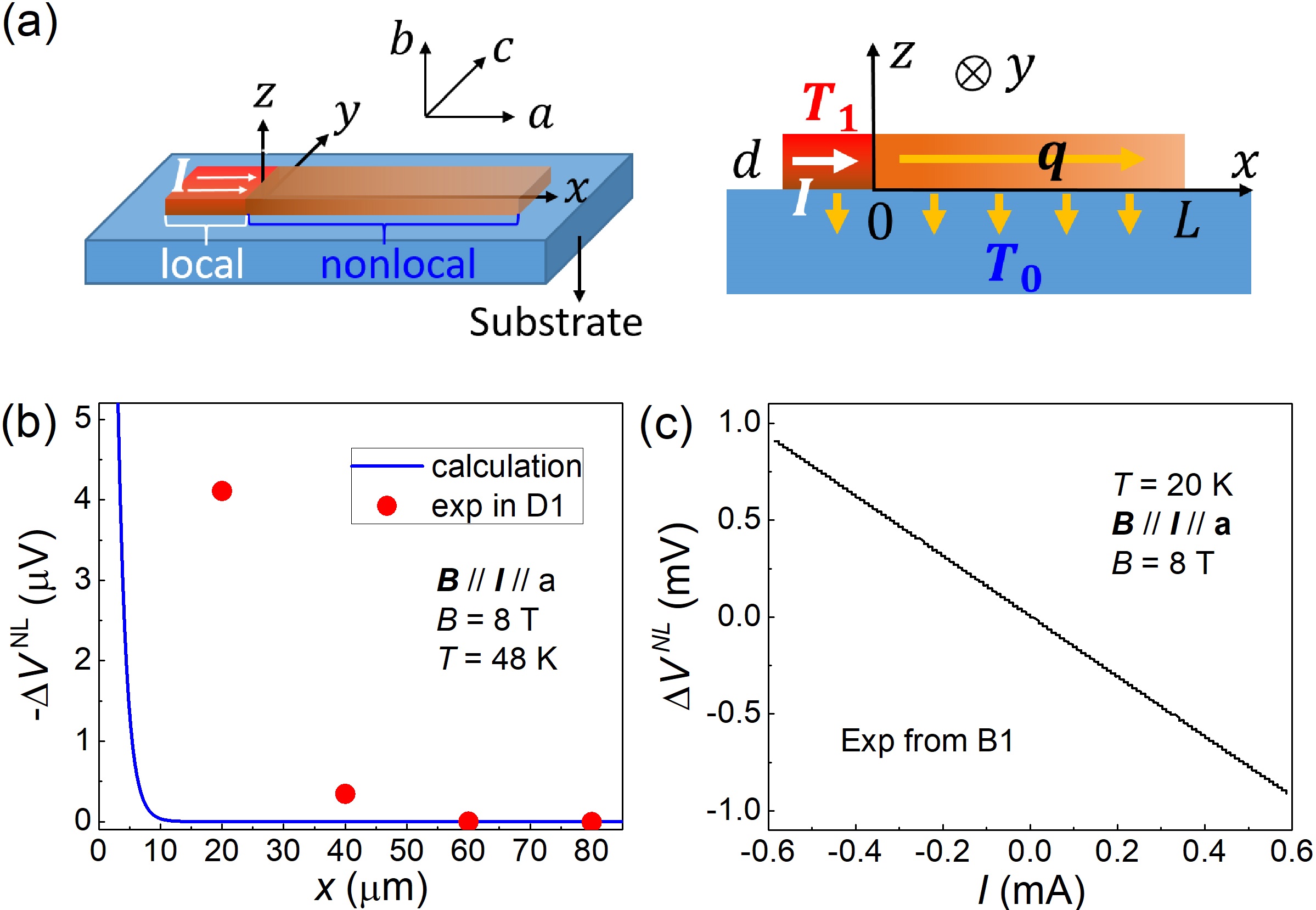}
	\caption{\textbf{Nonlocal transport induced by Joule heating.} (a) Left: schematic picture of the thermal diffusion in the sample, where the boundary between the local and nonlocal regions is the $yz$ plane at $x = 0$. Right: side view of the sample. The temperature at the top surface of the local region ($x < 0$) is $T_1$. The heat is drained through the bottom surface of the sample to the substrate at $T_0 < T_1$. (b) $x$-dependence of $-\Delta V^{\rm NL}$ obtained from the calculation [\eqref{B12}, blue line] and from the experiment at 8 T and 48 K for $B \parallel I \parallel a$-axis (red circles). (d) Linear relation between $I$ and $\Delta V^{\rm NL}$ measured in sample B1 with a DC technique. 
	}	\label{fig:Joule heating}
\end{figure}

To analyze the spurious nonlocal transport induced by Joule heating, we first make a crude hypothesis that the electric current applied in the local region ($x < 0$) causes a significant temperature rise (to $T_1$) at the top surface of the local area due to the Joule heating and the heat is drained through the bottom surface ($ac$ plane) of the sample which maintains a good thermal contact with the substrate held at a fixed cold temperature $T_0$, such that the bottom-surface temperature of the sample is always $T_0$ 
[Fig. \ref{fig:Joule heating}(a)] (more realistic simulations are presented later). The thermal diffusion in the nonlocal region ($x > 0$) causes a temperature gradient along the $x$-axis at the top surface, which leads to a finite voltage gradient due to the Seebeck effect. Here, the heat flows along both $x$ and $z$ axes and we assume $\partial T / \partial y = 0$.

The thermal diffusion process can be described in general by 
\begin{equation} 
\rho c_p \frac{\partial T}{\partial t} = \nabla \cdot (k \nabla T) + Q, \label{B1} 
\end{equation}
where $\rho$, $c_p$, $k$ and $Q$ are the material density, specific heat, thermal conductivity, and internal heat source (e.g., chemical reactions), respectively. For the steady state ($\partial T / \partial t = 0$) and without internal heat source ($Q = 0$), \eqref{B1} reduces to $\nabla \cdot (k \nabla T) = 0$, which can be further simplified with $\partial T / \partial y = 0$ to
\begin{equation}
\begin{split}
k_{xx} \frac{\partial^2 T}{\partial x^2} + k_{xz} \frac{\partial^2 T}{\partial x \partial z} + k_{zx} \frac{\partial^2 T}{\partial z \partial x} + k_{zz} \frac{\partial^2 T}{\partial z^2} = 0 \,\,.
\label{B2}
\end{split}
\end{equation}
Considering $k_{xz} = - k_{zx}$ under magnetic field, \eqref{B2} reduces to
\begin{equation}
\begin{split}
\alpha \frac{\partial^2 T}{\partial x^2} +  \frac{\partial^2 T}{\partial z^2} = 0 \,\,,
\label{B3}
\end{split}
\end{equation}
where $\alpha = k_{xx}/k_{zz}$. Here we can see that \eqref{B3} is the 2D anisotropic Laplace equation, so we only consider the $xz$-plane and specify the coordinate with $(x,z)$. For the configuration in Fig. \ref{fig:Joule heating}(a), we have the boundary conditions $k_{zz} \cdot \partial T(x,z) / \partial z |_{z = d} = I^2 R/w \cdot \delta(x)$ (Joule heating acts as a heat source at $(0,d)$ to maintain the temperature gradient, $d$ and $w$ are the sample thickness and width, respectively), $T(0 \le x \le +\infty,0) = T_0$ (temperature of the substrate), and $\partial T(x > 0,z) / \partial z |_{z = d} = 0$ (top surface is thermally insulated from the environment). The temperature distribution can be calculated by solving \eqref{B3}, yielding
\begin{equation}
\begin{split}
T(x, z) = T_0 + \frac{2 I^2 R}{\pi w \sqrt{k_{xx} k_{zz}}} \int_{0}^{\infty} \frac{\sinh(\sqrt{\alpha} k z)}{k \cosh(\sqrt{\alpha} k d)} \cos(k x) \, dk \,\,,
\label{B4}
\end{split}
\end{equation}
where $x \ge 0$ and $0 \le z \le d$. Since the nonlocal voltage was measured on the top surface, we only focus on the results at $z = d$, where \eqref{B4} approximately reduces to
\begin{equation}
\begin{split}
T(x, d) \approx T_0 + \frac{4 I^2 R}{\pi w \sqrt{k_{xx} k_{zz}}} \cdot e^{-\pi x/(2 d \sqrt{\alpha})}  \,\,,
\label{B6}
\end{split}
\end{equation}
where $x \gg d \sqrt{\alpha}$ is considered ($x \ge$ 20 $\mu$m and $d \approx$ 0.38 $\mu$m in sample D1, $\alpha$ is assumed to be 33, the same as the resistivity anisotropy). One can see that the temperature on the top surface decays exponentially along the $x$-axis.  

The temperature gradient at $z = d$ induces voltage gradient due to the thermoelectric effect
\begin{equation}
\begin{split}
\frac{\partial V(x,d)}{\partial x} = - S_{xx} \frac{\partial T(x,d)}{\partial x} \,\,,
\label{B8}
\end{split}
\end{equation}
where $S_{xx}$ is the Seebeck coefficients along the $x$-axes.
We define the nonlocal voltage as
\begin{equation} 
\Delta V^{\rm NL} (x) = V(x,d) - V(+ \infty,d), \label{B10} 
\end{equation}
which can be calculated with \eqref{B6} to be 
\begin{equation} 
\Delta V^{\rm NL}(x) \approx -\frac{4 I^2 R S_{xx}}{\pi w \sqrt{k_{xx} k_{zz}}} \cdot e^{-\pi x/(2 d \sqrt{\alpha})}. \label{B12} 
\end{equation}
Hence, one can see that the induced nonlocal voltage decays exponentially with $x$, characterized by the decay length $\lambda = 2 d \sqrt{\alpha} / \pi$. Therefore, in the case of the nonlocal trasport due to Joule heating, $\lambda$ scales linearly with the sample thickness $d$, but remains independent of the sample width, which is inconsistent with our experimental result. Also, by using $\alpha \approx 33$ and $d = 0.38$ $\mu$m for sample D1, we obtain $\lambda \approx$ 1.4 $\mu$m, which is much shorter than the experimental results of 8 $\mu$m [see Fig. \ref{fig:Joule heating}(b) for comparison].

More importantly, since the temperature gradient is sustained by the Joule heating ($I^2R$), the nonlocal voltage should present a quadratic dependence on the current, i.e. $\Delta V^{\rm NL} \propto I^2$, in the thermal scenario. We observed that $\Delta V^{\rm NL}$ always increases linearly with $I$, as shown exemplarily in Fig. \ref{fig:Joule heating}(d) for sample B1. This ohmic relation between $\Delta V^{\rm NL}$ and $I$ gives the most convincing evidence against the thermal scenario for the nonlocal transport.

\subsection{3D simulation of thermal diffusion induced by Joule heating in sample D1} 

\begin{figure}[h]
	\centering
	\includegraphics[width=10 cm]{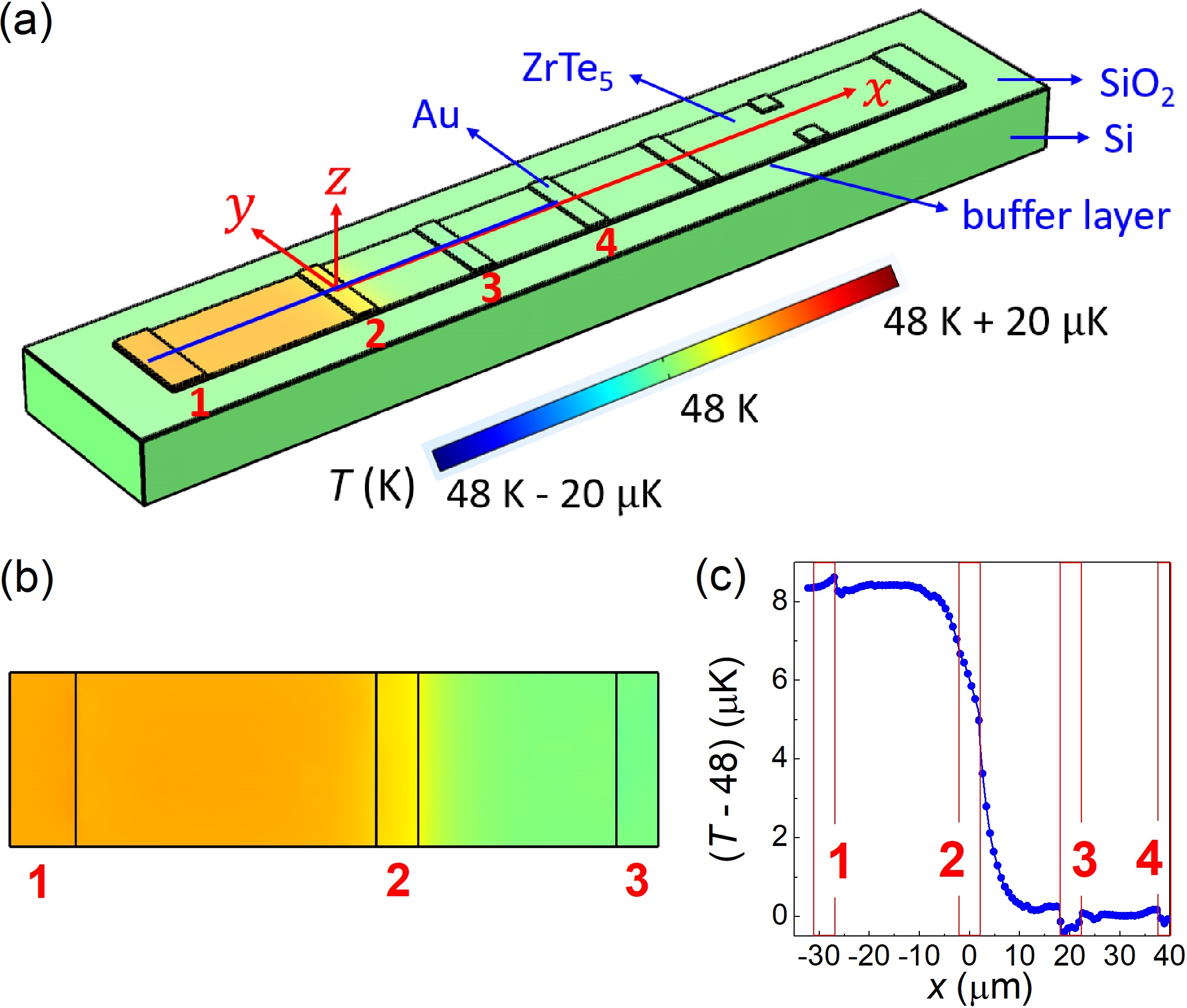}
	\caption{\textbf{3D simulation of thermal diffusion induced by Joule heating in sample D1.} (a) 3D view of the simulated system. The local region of the sample between electrodes 1 and 2 acts as a heat source due to Joule heating. The heat is primarily drained through the bottom surface of the substrate held at a fixed cold temperature of 48 K. To consider a worse-case scenario of a very poor thermal contact between the sample and the substrate, a 100-nm-thick buffer layer with a low thermal conductivity (0.1 Wm$^{-1}$K$^{-1}$) was incorporated in the model. (b) Temperature distribution on the sample surface in the region between electrodes 1 and 3. (c) $x$-dependence of ($T - 48$ K) (relative temperature to 48 K). The position of the electrodes are marked.
	}
	\label{fig:thermal diffusion for D1}
\end{figure}

Thermal-diffusion simulations were performed for a system consisting of sample D1, a buffer layer providing a thermal resistance between the sample and the substrate, and the substrate (doped Si coated with SiO$_2$), by considering the Joule heating in the local region of the sample [Fig. \ref{fig:thermal diffusion for D1}]. The 3D boundary condition for the simualtions was set by the measurement configuration of sample D1. The Joule heating in the local part between electrodes 1 and 2 acts as a heat source. The heat is primarily drained through the bottom surface of the substrate, which is held at a fixed cold temperature of 48 K. A 100-nm-thick buffer layer with a very low thermal conductivity (0.1 Wm$^{-1}$K$^{-1}$) was considered to model a poor thermal contact between the sample and the substrate.

In the steady state, the temperature distribution can be calculated with the inputs of Joule heating power $Q$ and thermal conductivity $k$. The heat power density $p$ generated by Joule heating is given by $p = j^2 \rho$, where $j$ and $\rho$ represent the current density and the resistivity of the sample, respectively. For sample D1, this yields $p \approx 1 \times 10^7$ Wm$^{-3}$ at 48 K and 8 T for $I$ = 4 $\mu$A in the $B \parallel I \parallel a$-axis configuration.

For the Te-flux-grown ZrTe$_5$ crystals, the reported thermal conductivity along the $a$-axis ($k_a$) is in the range 10 -- 100 Wm$^{-1}$K$^{-1}$ at 48 K \cite{Feng2020}. We performed the simulations with $k_a$ = 10 Wm$^{-1}$K$^{-1}$, to consider the worst-case scenario. The thermal conductivity has two parts, phononic and electronic contributions. The electronic contribution is negligible due to the very low electrical conductivity, and the anisotropy of the lattice contribution is reported to be $k_a^{ph} : k_c^{ph} : k_b^{ph} \approx$ 10 : 5 : 1 \cite{Feng2020}. Thus, we use $k_a$ = 10 Wm$^{-1}$K$^{-1}$, $k_b$ = 5 Wm$^{-1}$K$^{-1}$, and $k_c$ = 1 Wm$^{-1}$K$^{-1}$ as the input for our simulations. For the gold film used for making the electrodes, the thermal conductivity is estimated to be 15 Wm$^{-1}$K$^{-1}$ at 48 K \cite{Zink2005}. The substrate is SiO$_2$ 290 nm on doped Si (B doped, 0.001 -- 0.005 $\Omega$cm). For the doped Si wafer, the thermal conductivity at 48 K is around 100 Wm$^{-1}$K$^{-1}$ \cite{Li2017}. For the SiO$_2$ layer, the thermal conductivity is around 0.3 Wm$^{-1}$K$^{-1}$ at 48 K \cite{Li2017}. For the buffer layer to model a bad contact between sample and substrate, we assumed only 0.1 Wm$^{-1}$K$^{-1}$, as already mentioned.

The simulated temperature distribution shows that the temperature rise due to Joule heating is only less than 9 $\mu$K [Fig. \ref{fig:thermal diffusion for D1}(c)] even in this worst-case scenario. Therefore, the temperature increase due to the current (and the associated thermoelectric effect) was negligible in sample D1.

\section{2D diffusion model}

The diffusion equation describes how particles, energy, or other quantities spread out in space over time. For a two-dimensional (2D) steady state, the diffusion in an anisotropic medium can be described by the anisotropic Laplace equation
\begin{equation} 
\alpha \frac{\partial^2 n}{\partial x^2} +  \frac{\partial^2 n}{\partial y^2} = 0,  \label{D1} 
\end{equation}
where $\alpha = D_{x}/D_{y}$ is the anisotropy of the diffusion coefficients and $n$ represents the particle density.

\begin{figure}[h]
	\centering
	\includegraphics[width=7 cm]{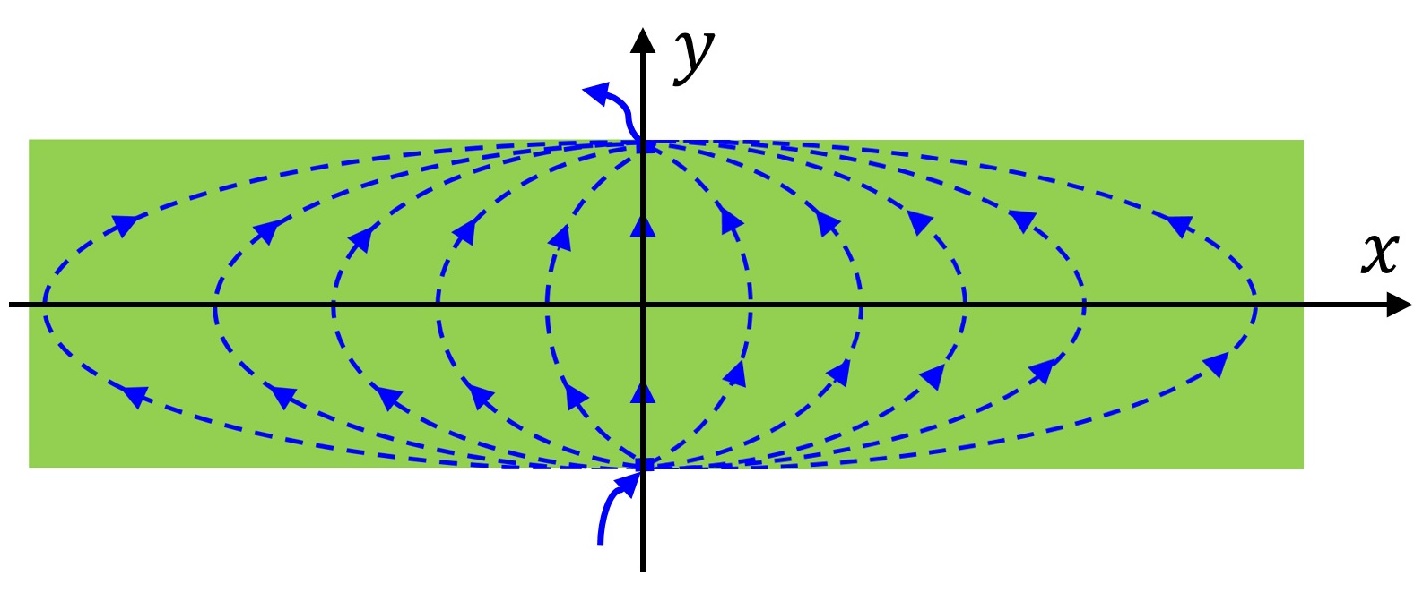}
	\caption{\textbf{2D diffusion for point source/sink.} Illustration of the particle diffusion in a 2D strip between a point source and a point sink on the edges. The particle flow is vertically injected from $(0,-w/2)$ and vertically drained through $(0,w/2)$.
	}	\label{fig:2D diffusion_1}
\end{figure}

Let us consider an infinitely long 2D strip of width $w$ ($-\infty < x < +\infty$, $-w/2 \le y \le w/2$) and put a point source at $(0, -w/2)$ and a point sink at $(0, w/2)$, which gives the boundary condition $- D_{y} \frac{\partial n}{\partial y} |_{y= \pm w/2} = I_D \delta(x)$ (see Fig. \ref{fig:2D diffusion_1}). Since the particle density vanishes at far ends of the strip, we have $n(x=\pm \infty,y) = 0$. The remaining boundaries are insulated. With this boundary condition, the solution of \eqref{D1} becomes
\begin{equation}
\begin{split}
n(x, y) = - \frac{2 I_D}{\pi \sqrt{D_{x} D_{y}}} \int_{0}^{\infty} \frac{\sinh(\sqrt{\alpha} k y)}{k \cosh(\sqrt{\alpha} k w/2)} \cos(k x) \, dk \,\,.
\label{D2}
\end{split}
\end{equation}
The particle density difference between the two edges at the position $x$, $\Delta n(x) \equiv n(x,-w/2) - n(x,w/2)$, can be calculated as
\begin{equation} 
\Delta n (x) \approx  \frac{4 I_D}{\pi \sqrt{D_{x} D_{y}}} \cdot e^{-\pi x/(2w\sqrt{\alpha})}  \,\,\,\, (x \gg w),  \label{D3} 
\end{equation}
which indicates that the particle density difference decays exponentially with $x$, characterized by a decay length $\lambda = 2 w \sqrt{\alpha} / \pi$. If the diffusion current of the particles generates a transverse electric field as in the case of the inverse spin Hall effect \cite{Sinova2015}, it results in the voltage gradient in the $x$ direction. Hence, in this scenario, \eqref{D3} would lead to 
\begin{equation} 
\Delta V (x) \propto e^{-\pi x/(2w\sqrt{\alpha})}  \,\,\,\, (x \gg w).  \label{D4} 
\end{equation}
This shows that the induced voltage decays exponentially with $x$, characterized by a decay length $\lambda = 2 w \sqrt{\alpha} / \pi$ that increases linearly with the width of the strip. This is similar to what we observed in ZrTe$_5$.

\begin{figure}[h]
	\centering
	\includegraphics[width=11 cm]{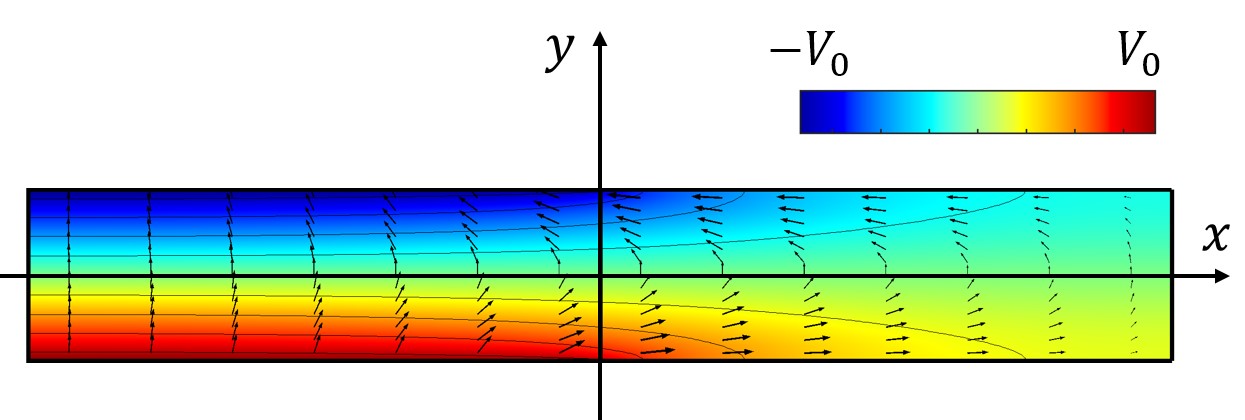}
	\caption{\textbf{2D diffusion for line source/sink.} Schematic of the particle diffusion in a 2D strip between a line source and a line sink on the edges. The particle flow is injected from $(x<0,-w/2)$ and drained through $(x<0,w/2)$.
	}	\label{fig:2D diffusion_2}
\end{figure}

If the electric current in the local region in our experiment is producing a polarization of some degree of freedom at the opposing edges of the sample as in the case of the spin Hall effect \cite{Sinova2015}, the boundary condition of the 2D diffusion problem would be better approximated by considering a line source at $(x<0, -w/2)$ and a line sink at $(x<0, w/2)$. The solution of \eqref{D1} in this case is schematically shown in Fig. \ref{fig:2D diffusion_2}. It is written mathematically as
\begin{equation} 
n(x,y) = -\frac{2 j_D w}{\pi^2 \sqrt{D_x D_y}} \sum_{n=0}^\infty \frac{(-1)^n}{(2n+1)^2} \cdot e^{-\frac{(2n+1)\pi x}{\sqrt{\alpha}w}} \sin\left(\frac{(2n+1)\pi y}{w}\right)  \,\,\,\, (x > 0),  \label{D5} 
\end{equation}
\begin{equation} 
n(x,y) = -\frac{j_D}{D_y}y + \frac{2 j_D w}{\pi^2 \sqrt{D_x D_y}} \sum_{n=0}^\infty \frac{(-1)^n}{(2n+1)^2} \cdot e^{\frac{(2n+1)\pi x}{\sqrt{\alpha}w}} \sin\left(\frac{(2n+1)\pi y}{w} \right)\,\,\,\, (x \le 0),   \label{D6} 
\end{equation}
where the boundary condition for the particle current was replaced with $- D_{y} \frac{\partial n}{\partial y} |_{y= \pm w/2} = j_D \theta(-x)$ (where $j_D$ is the current density of the particles flow for $x \ll 0$ and $\theta(x)$ is the step function). For $|x| \gg w$, the dominant contribution to \eqref{D5} and \eqref{D6} comes from the $n = 0$ term, which leads to
\begin{equation} 
n(x,y) \approx -\frac{2 j_D w}{\pi^2 \sqrt{D_x D_y}} \cdot e^{-\pi x / (w \sqrt{\alpha})} \sin\left(\frac{\pi y}{w}\right)  \,\,\,\, (x \gg w),  \label{D7} 
\end{equation}
\begin{equation} 
n(x,y) \approx -\frac{j_D}{D_y} y \,\,\,\, (x \ll -w).  \label{D8} 
\end{equation}
This leads to
\begin{equation} 
\Delta n (x) \propto e^{-\pi x/(w\sqrt{\alpha})}  \,\,\,\, (x \gg w),  \label{D9} 
\end{equation}
Hence, the resulting $\Delta V (x)$ behaves essentially the same as the point-source case, but now with $\lambda = w \sqrt{\alpha} / \pi$.

\section{Measurements in the standard nonlocal configuration}

Most nonlocal transport experiments performed in the past \cite{Abanin2011, Gorbachev2014, Sui2015, Shimazaki2015, CZhang2017} used a strip-shaped sample with pairs of contacts placed at the opposing edges of the strip. Such a contact arrangement is usually used as a ``Hall bar", in which current flows along the strip and the Hall voltage is detected using an opposing pair of contacts. In the nonlocal measurements, however, current is injected from one of the electrodes of a pair and drained through the other, which results in the current to mainly flow transverse to the strip. The nonlocal voltage is detected as a voltage drop occurring in other pairs of contacts, that are placed at some distance from the current injection/drain pair. We call it ``standard nonlocal configuration" in this work. We performed additional experiments in this standard configuration using sample B4 (for $I \parallel c$, Fig. \ref{fig:standard configuration_1}) and sample B5 (for $I \parallel b$, Fig. \ref{fig:standard configuration_2}). 

\subsection{Sample B4 for $I \parallel c$}

\begin{figure}[h]
	\centering
	\includegraphics[width=12 cm]{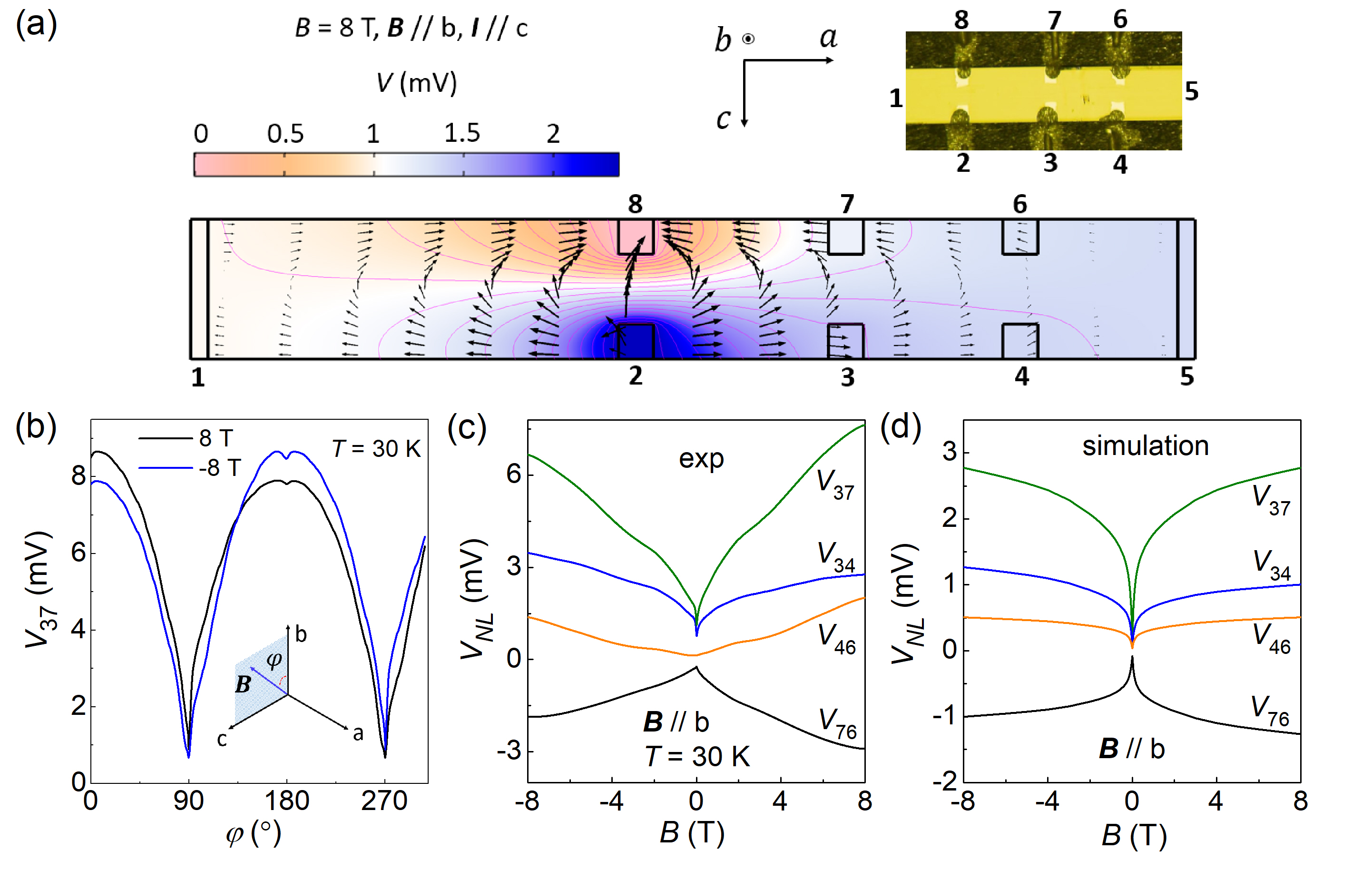}
	\caption{\textbf{Standard-configuration measurements in sample B4.} (a) Upper-right inset shows the photograph of sample B4 with contact arrangement. The main panel shows the simulated current and voltage distributions in the $ac$ plane at 30 K and 8 T for $B \parallel b$-axis. Color scale represents the voltage value, while arrows represent the current density and direction. (b) Magnetic-field-orientation dependence of $V_{37}$ at 30 K for $\pm$8 T rotated in the $bc$ plane. Inset shows the definition of the magnetic-field orientation $\varphi$. (c),(d) $B$-dependences of the nonlocal voltage at various pairs of electrodes at 30 K for $B \parallel b$-axis, obtained from experiments (c) and simulations (d).       
	}	\label{fig:standard configuration_1}
\end{figure}

In sample B4, three pairs of opposing side contacts were made on the side faces of a thin rectangular bulk sample [see Fig. \ref{fig:standard configuration_1}(a)]. The electric current was injected from electrode 2 and drained through grounded electrode 8, such that it primarily flows along the $c$-axis. The 2D simulation of the current and voltage distributions obeying the Ohm's law performed for the $ac$ plane is shown in Fig. \ref{fig:standard configuration_1}(a). One can see that the current and the voltage gradients decay rather slowly with the distance from the electrodes 2 and 8. This is due to the strong resistivity anisotropy which causes the current to preferentially flow along the $a$-axis. Consequently, this measurement configuration is not suitable for separating the unusual nonlocal transport component from the Ohm's-law contribution.
In fact, the observed nonlocal voltage appears to be dominated by the Ohm's-law contribution for the following reasons: (i) The magnetic-field-orientation dependence of the nonlocal voltage $V_{37}$ (measured between electrodes 3 and 7) is similar to that of the local magnetoresistance \cite{Wang2022} without showing a sharp peak for $B \parallel I$ [Fig. \ref{fig:standard configuration_1}(b)]. (ii) The $B$-dependences of the nonlocal voltages obtained by the 2D simulation [Fig. \ref{fig:standard configuration_1}(d)] roughly matches the experimental results [Fig. \ref{fig:standard configuration_1}(c)].


\subsection{Sample B5 for $I \parallel b$}

\begin{figure}[h]
	\centering
	\includegraphics[width=11 cm]{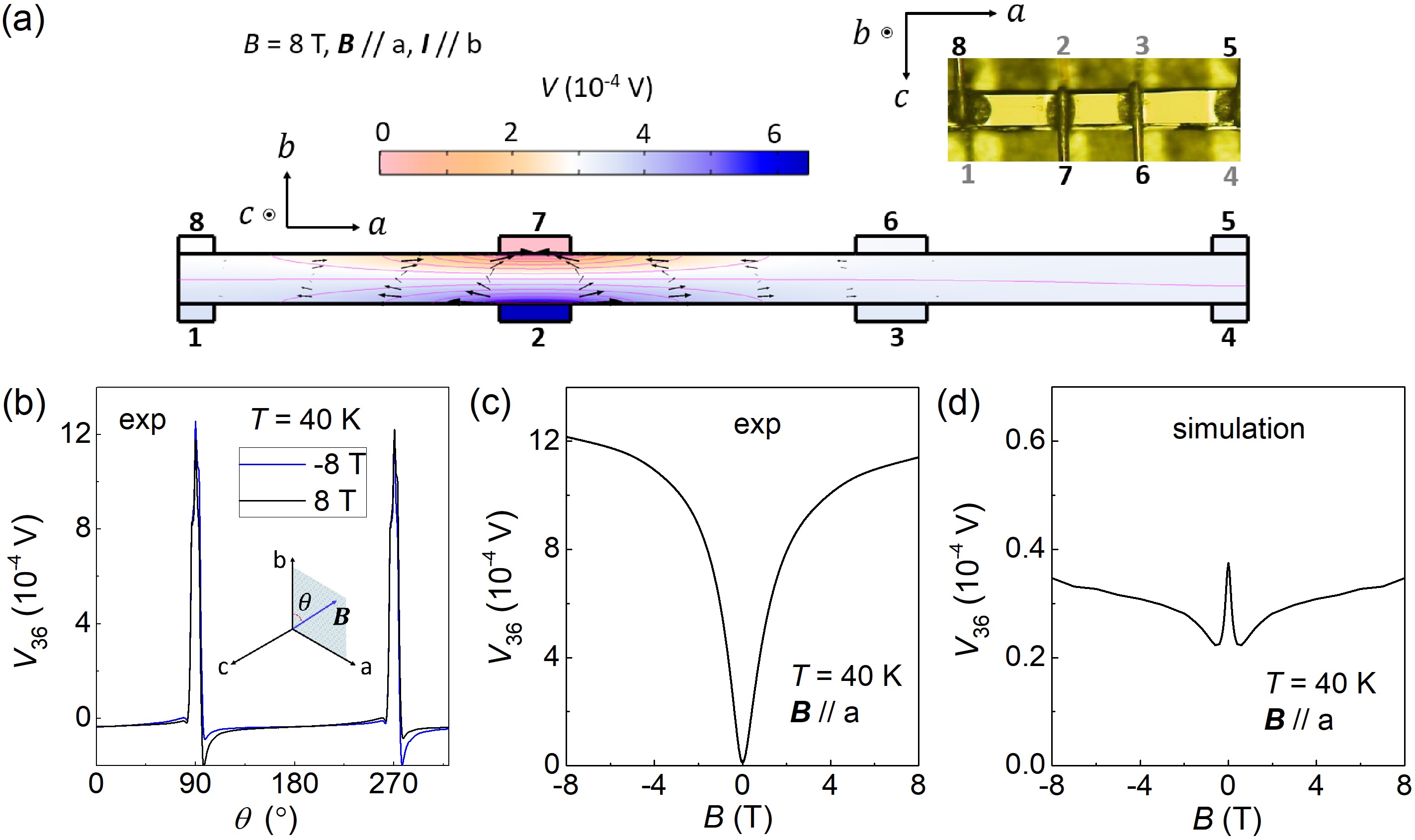}
    \caption{\textbf{Standard-configuration measurements in sample B5.} (a) Upper-right inset shows the photograph of sample B5 with contact arrangement (electrodes 1--4 were made on the bottom surface and are invisilbe in this photograph). The main panel shows the simulated current and voltage distributions in the $ab$ plane at 40 K and 8 T for $B \parallel a$-axis. Color scale represents the voltage value, while arrows represent the current density and direction. (b) Magnetic-field-orientation dependence of $V_{36}$ at 30 K for $\pm$8 T rotated in the $ab$ plane. Inset shows the definition of the magnetic-field orientation $\theta$. (c),(d) $B$-dependences of the nonlocal voltage $V_{36}$ at 40 K for $B \parallel a$-axis, obtained from experiment (c) and simulation (d).      
	}
	\label{fig:standard configuration_2}
\end{figure}

In sample B5, four pairs of contacts were made on the top and bottom surfaces ($ac$ plane) of a rectangular bulk sample [see Fig. \ref{fig:standard configuration_2}(a)]. The electric current was injected from electrode 2 and drained through grounded electrode 7, such that it primarily flows along the $b$-axis. The 2D simulation of the current and voltage distributions obeying the Ohm's law performed for the $ab$ plane is shown in Fig. \ref{fig:standard configuration_2}(a). One can see that the current and the voltage gradients decay relatively quickly with distance from the electrodes 2 and 7, due to a small thickness of the sample. This limited thickness prevents long-range charge carrier diffusion, resulting in negligible nonlocal voltage at electrodes 3 and 6 due to Ohm's-law contribution. The magnetic-field-orientation dependence of the nonlocal voltage $V_{36}$ (between electrodes 3 and 6) presents a sharp peak when $\theta$ is close to 90$^{\circ}$ (i.e. $B \parallel a$-axis) as shown in Fig. \ref{fig:standard configuration_2}(b), which is similar to the results in Fig. 1b of the main text. 
The sensitivity of the nonlocal signal on $B \parallel a$-axis and the strange symmetry breaking near $B \parallel a$-axis seem to indicate that the same unusual physics is playing a role in this configuration. 
The $B$-dependence of the nonlocal voltage $V_{36}$ (= $V_3 - V_6$) [Fig. \ref{fig:standard configuration_2}(c)] is totally different from the result of the Ohm's-law simulation [Fig. \ref{fig:standard configuration_2}(d)], which also supports the conclusion that the trivial contribution is negligible in this configuration.
Importantly, the data in Fig. \ref{fig:standard configuration_2}(b) seem to suggest that $B \parallel a$ is crucial for the unusual nonlocal transport but $B \parallel I$ is not.

\section{Measurements of a samples with $T_p$ = 108 K}

\begin{figure}[h]
	\centering
	\includegraphics[width=11 cm]{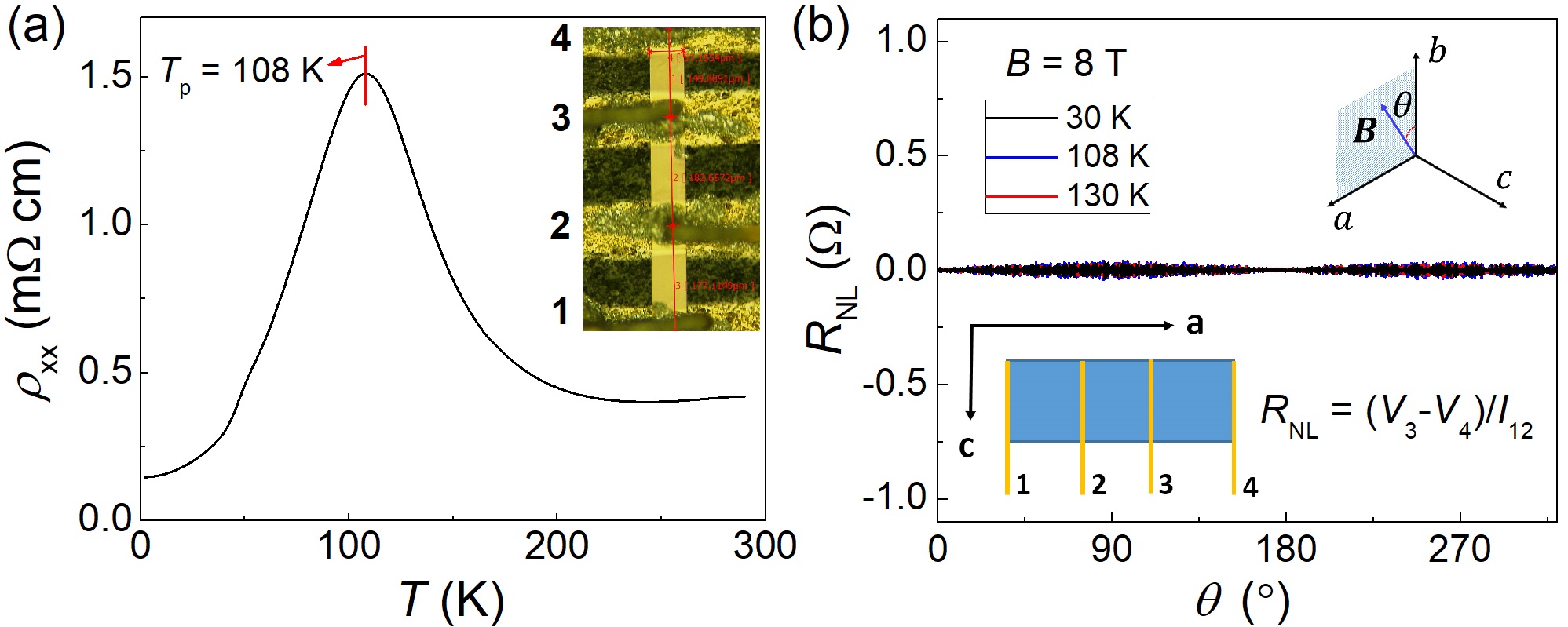}
	\caption{\textbf{Nonlocal transport measurements in sample B6 with $T_p$ = 108 K.} (a) Temperature dependence of $\rho_{xx}$. Inset shows the sample photograph with contact arrangement. (b) Magnetic-field-orientation dependence of the nonlocal resistance at 30, 108, and 130 K in 8 T rotated in the $ab$ plane, presenting only noise. Insets show the measurement configuration and the definition of the magnetic-field orientation $\theta$. 
	}
	\label{fig:high Tp}
\end{figure}

The transport properties of ZrTe$_5$ samples showing a high $T_p$ value are very different from those of $T_p \approx$ 0 K samples \cite{Wang2022, Wang2023, Wang2025}. To gain additional insight into the origin of the unusual nonlocal transport in ZrTe$_5$, we investigated the nonlocal transport in a ZrTe$_5$ sample with $T_p$ = 108 K [Fig. \ref{fig:high Tp}(a)]. The magnetic-field-orientation dependence of the nonlocal resistance for $B$ rotated in the $ab$ plane was measured in three regimes: the strong topological insulator (TI) phase (30 K), near the Lifshitz transition point (108 K), and the weak TI phase (130 K) \cite{Xu2018}. In all three regimes, the nonlocal signal was vanishingly small at any angle $\theta$ [Fig. \ref{fig:high Tp}(b)]. This result shows the absence of unusual nonlocal transport in samples with a high $T_p$.  

\newpage

\section{Reproducibility check in sample D3}

To check for the reproducibility of the observed nonlocal transport phenomenon including the very unusual nonlocal Hall effect, we performed similar experiments on sample D3 whose width and thickness were 14.6 $\mu$m and 0.36 $\mu$m, respectively. The connection to electrode 4 was broken during the experiment, which made it impossible to measure the decay length of the nonlocality in this sample. Nevertheless, the voltage differences $V_3 - V_8$ and $V_6 - V_7$ were measured without problem, which allowed us to confirm the reproducibility of most of the unusual aspects of the nonlocal transport.

\subsection{Nonlocal resistance}

\begin{figure}[h]
	\centering
	\includegraphics[width=10 cm]{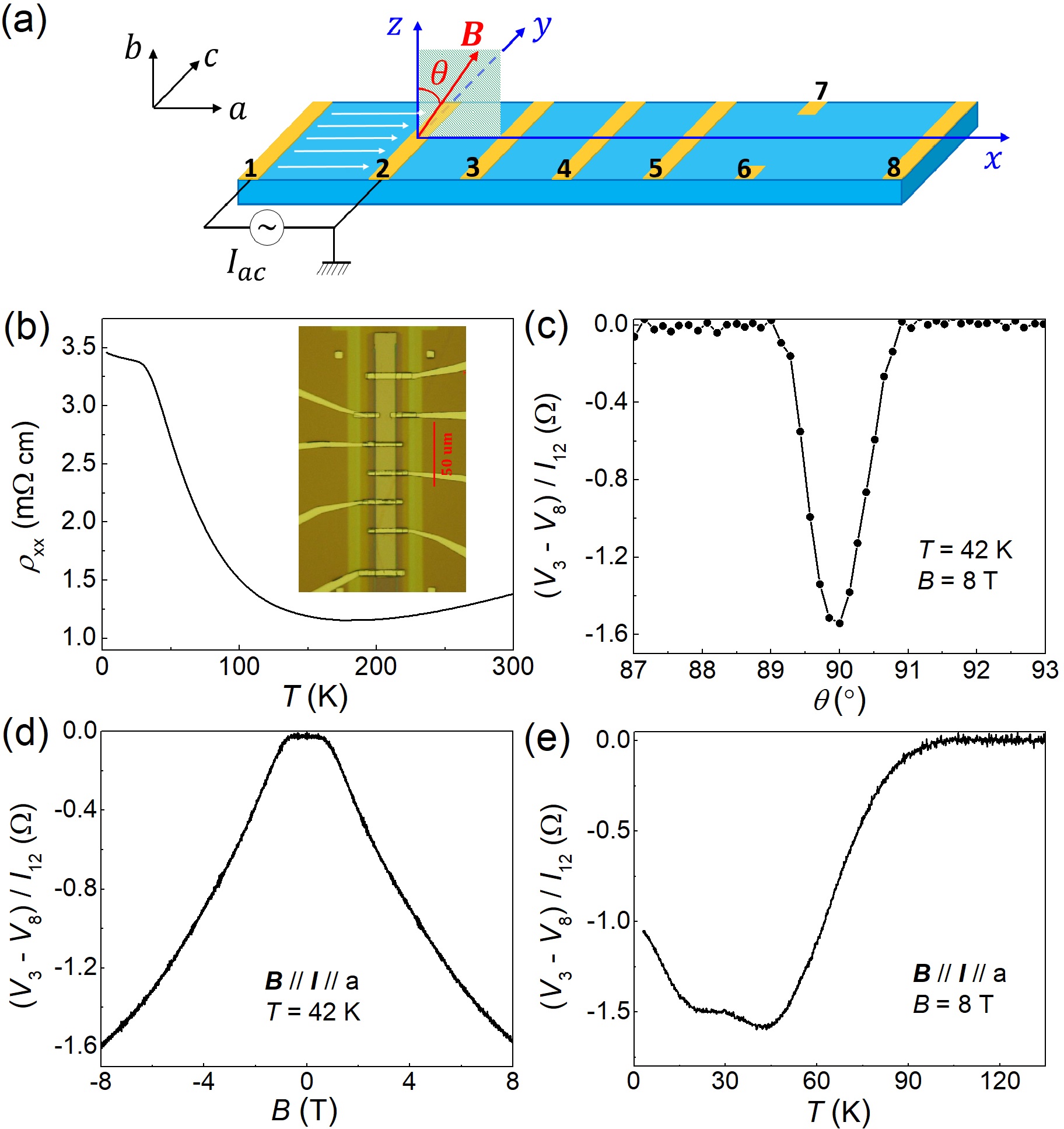}
	\caption{\textbf{Nonlocal resistance $R_3^{\rm NL}$ in sample D3.} (a) Schematic of the measurement configuration. Current $I_{12}$ was injected from electrode 1 and drained through the grounded electrode 2. The magnetic field $B$ was applied in the $xz$ plane with an angle $\theta$ measured from the $z$-axis. The $xyz$ transport axes correspond to $acb$ crystal axes. The $x$ coordinate is measured from the center of electrode 2. (b) Temperature dependence of $\rho_{xx}$. The inset shows the photograph of sample D3. (c) $\theta$-dependence of $(V_3 - V_8)/I_{12}$ in 8 T at 42 K. (d) $B$-dependence of $(V_3 - V_8)/I_{12}$ measured in $B \parallel I \parallel a$-axis ($\theta$ = 90$^\circ$) at 42 K. (e) Temperature dependence of $(V_3 - V_8)/I_{12}$ at 8 T for $B \parallel I \parallel a$-axis.
	}
	\label{fig:nonlocal resistance for D3}
\end{figure}

The behavior of the nonlocal resistance $R_n^{\rm NL}$ with $n$ = 3 is shown in Fig. \ref{fig:nonlocal resistance for D3}. Three unusual features, (i) $R_3^{\rm NL}$ appears only when the magnetic-field angle $\theta$ was within $\pm$1$^{\circ}$ from 90$^{\circ}$, (ii) There is a threshold value of $B$ for the appearance of $R_3^{\rm NL}$, (iii) $R_3^{\rm NL}$ presents a non-monotonic temperature dependence and persists up to $\sim$100 K, were all reproduced.

\subsection{Nonlocal Hall effect}

\begin{figure}[h]
	\centering
	\includegraphics[width=6 cm]{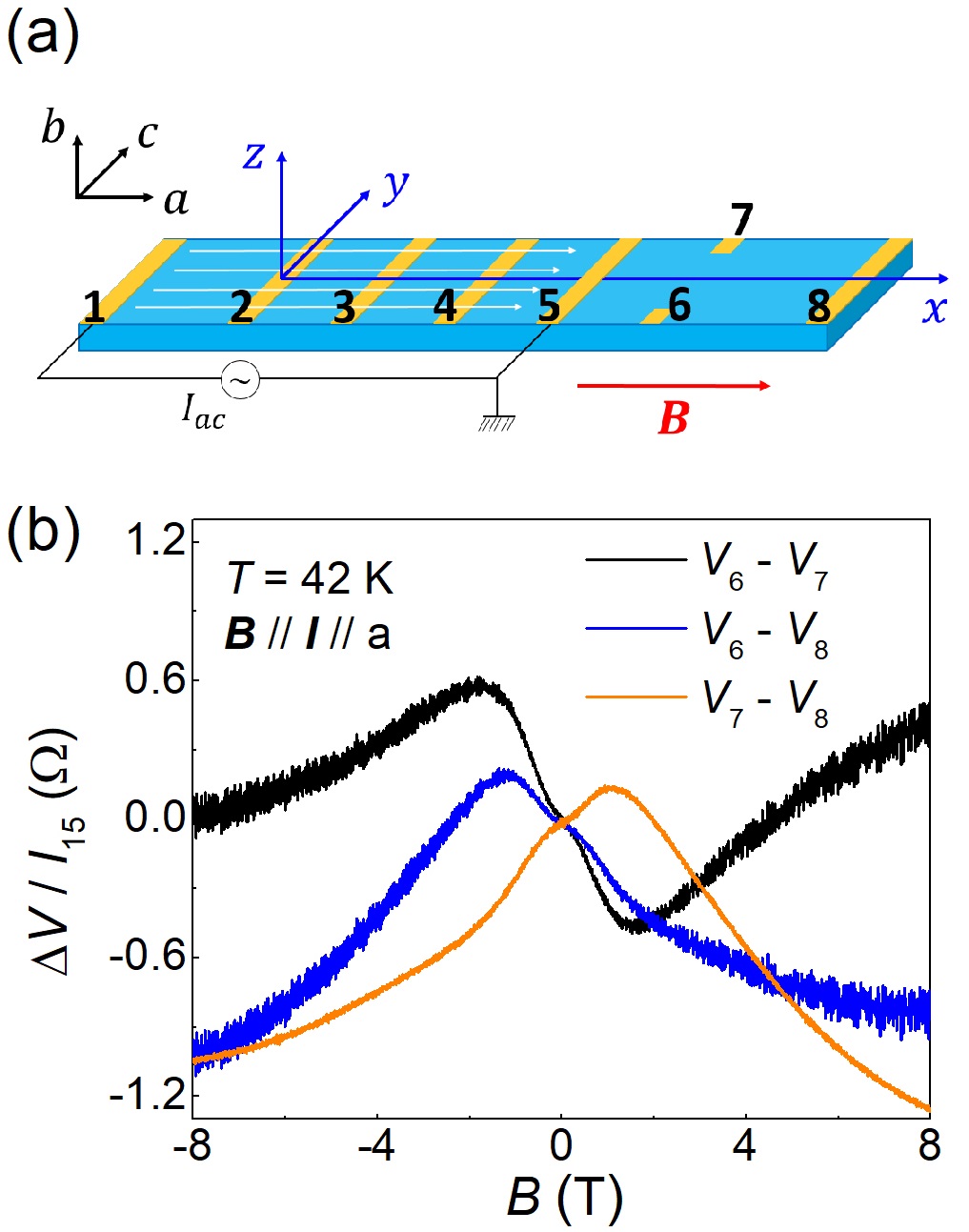}
	\caption{\textbf{Nonlocal Hall effect in sample D3.} (a) Measurment configuration for the current $I_{15}$. (b) $B$-dependence of $(V_6-V_7)/I_{15}$, $(V_6-V_8)/I_{15}$, and $(V_7-V_8)/I_{15}$ (all without antisymmetrization).
	}
	\label{fig:nonlocal Hall for D3}
\end{figure}

The unusual nonlocal Hall effect was also reproduced for $I_{15}$ in sample D3 at 42 K as shown in Fig. \ref{fig:nonlocal Hall for D3}.

\newpage

\bibliography{ZrTe5_bibliography}